\documentclass[acmlarge,screen,nonacm]{acmart}

\usepackage{enumitem}
\usepackage{multirow}
\usepackage{tabu}
\usepackage[utf8]{inputenc}
\usepackage[T1]{fontenc}
\usepackage[english]{babel}
\usepackage{tablefootnote}
\usepackage{longtable}
\usepackage[most]{tcolorbox}
\usepackage{booktabs, caption, makecell}
\usepackage{threeparttable}
\usepackage{fixltx2e}%
\usepackage{hyperref}
\usepackage{array}
\hypersetup{colorlinks=true, urlcolor=blue}

\AtBeginDocument{%
  \providecommand\BibTeX{{%
    \normalfont B\kern-0.5em{\scshape i\kern-0.25em b}\kern-0.8em\TeX}}}

\setcopyright{none}
\copyrightyear{2023}
\acmYear{2023}
\acmDOI{XXXXXXX.XXXXXXX}

\setcopyright{none}
\renewcommand\footnotetextcopyrightpermission[1]{}
\newcommand{\rqone}{\textbf{RQ\textsubscript{1}.}What is the trend of the number of studies on build?}
\newcommand{\rqtwo}{\textbf{RQ\textsubscript{2}.} What are the main build-related topics covered by the literature?}
\newcommand{\rqthree}{\textbf{RQ\textsubscript{3}.} What are the existing techniques for addressing long build duration?}
\newcommand{\rqfour}{\textbf{RQ\textsubscript{4}.} What are the existing techniques for build quality improvement?}
\newcommand{\rqfive}{\textbf{RQ\textsubscript{5}.} Which metrics are used for build studies?}
\newcommand{\rqsix}{\textbf{RQ\textsubscript{6}.} Which datasets are publicly available for build studies?}

\newenvironment{summary}[1][RQ]
{\begin{tcolorbox}[
    enhanced,
    attach boxed title to top left={xshift=1em,yshift=-\tcboxedtitleheight/2},
    colback=black!5!white,
    colframe=black!70!white,
    colbacktitle=red!80!black,
    drop shadow={black!50!white},
    coltitle=white,
    top=0.15in,
    title=Summary of #1,
    boxed title style={size=small,colback=gray}]}
{\end{tcolorbox}}

\begin{document}

\title{Build Optimization: A Systematic Literature Review}

\author{Henri A{\"i}dasso}
\orcid{0009-0004-1625-0159}
\affiliation{%
  \institution{\'Ecole de technologie sup\'erieure}
  \streetaddress{Software and IT Engineering Department}
  \city{Montreal}
  \state{Qu\'ebec}
  \country{Canada}}
\email{henri.aidasso.1@ens.etsmtl.ca}

\author{Mohammed Sayagh}
\orcid{0000-0002-2724-0034}
\affiliation{%
  \institution{\'Ecole de technologie sup\'erieure}
  \streetaddress{Software and IT Engineering Department}
  \city{Montreal}
  \state{Qu\'ebec}
  \country{Canada}}
\email{mohammed.sayagh@etsmtl.ca}

\author{Francis Bordeleau}
\orcid{0000-0001-7727-3902}
\affiliation{%
  \institution{\'Ecole de technologie sup\'erieure}
  \streetaddress{Software and IT Engineering Department}
  \city{Montreal}
  \state{Qu\'ebec}
  \country{Canada}}
\email{francis.bordeleau@etsmtl.ca}

\begin{abstract}
    Continuous Integration (CI) consists of an automated build process involving continuous compilation, testing, and packaging of the software system. While CI comes up with several advantages related to quality and time to delivery, CI also presents several challenges addressed by a large body of research. To better understand the literature so as to help practitioners find solutions for their problems and guide future research, we conduct a systematic review of 97 studies on build optimization published between 2006 and 2024, which we summarized according to their goals, methodologies, used datasets, and leveraged metrics. The identified build optimization studies focus on two main challenges: (1) long build durations, and (2) build failures. To meet the first challenge, existing studies have developed a range of techniques, including predicting build outcome and duration, selective build execution, and build acceleration using caching or repairing performance smells. The causes of build failures have been the subject of several studies, leading to the development of techniques for predicting build script maintenance and automating repair. Recent studies have also focused on predicting flaky build failures caused by environmental issues. The majority of these techniques use machine learning algorithms and leverage build metrics, which we classify into five categories. Additionally, we identify eight publicly available build datasets for build optimization research. 
     
\end{abstract}



\maketitle

\section{Introduction}
\label{intro}

The build is the pipeline (also known as CI pipeline) that is typically triggered automatically to convert the source code files into a standalone software artifact ready for end-users. It generally involves several steps such as compiling the source code files, running static analysis, executing test cases, and packaging the software \cite{zhang_buildsonic_2022, jin_which_2022}. In general, a newer build output contains more features and fewer bugs. With the adoption of DevOps practices in the technology industry, the build process has been automated through Continuous Integration (CI) pipelines. CI is triggered at each submission of changes by developers to the central version control system (VCS) \cite{shahin_continuous_2017}. Its execution produces log data that the developers use to diagnose errors in the event of build failures. 

CI has many benefits related to the acceleration of feedback and release cycles. On the one hand, it provides developers with early feedback on their code issues (e.g., bugs, code smells, quality smells, and merge conflicts) \cite{hilton_usage_2016}. This way, developers can quickly diagnose errors using logs and fix them immediately to minimize the presence of bugs in production \cite{brandt_logchunks_2020}. As a result, the quality of software products resulting from CI is significantly improved \cite{hilton_usage_2016}. On the other hand, projects that use CI have shorter and more frequent release cycles and tend to maintain their teams more productive \cite{shahin_continuous_2017}. Indeed, developers are encouraged to integrate their changes frequently, resulting in multiple software versions per day \cite{memon_taming_2017}. 

The extensive use of CI has raised several challenges mainly related to the increasingly long build times and build failures \cite{hilton_trade-offs_2017}. CI costs have been estimated at millions of dollars at Google and Mozilla \cite{zhang_buildsonic_2022}. These costs include the infrastructure and energy costs of running builds, as well as those associated with reduced developer productivity due to failures \cite{seo_programmers_2014, memon_taming_2017, olewicki_towards_2022}. To address these challenges, researchers have undertaken several studies and developed techniques for improving CI build, mainly by leveraging the build data generated during the CI build process \cite{jin_what_2021}.

In the aim of facilitating the usage and optimization of the CI-improving techniques, it is important to gain a good overview of the existing practices for build optimization. For example, \citet{jin_what_2021} surveyed the existing state-of-the-art build and test selection techniques. However, their study is scoped to build time reduction techniques and is limited to only 10 papers. To the extent of our knowledge, there is no systematic literature review that summarizes the existing practices in the literature regarding build optimization.

Hence, in this paper, we conduct a thorough systematic literature review (SLR) on build optimization. We obtained a total of 97 papers published between 2006 and 2024. The goal of our SLR is to help developers identify the approaches, techniques, and practices that exist for their problems and help researchers identify gaps in the literature on CI pipeline optimization. To do so, we address the following research questions (RQs): 

\begin{itemize}
    \item \textbf{\rqone}
    
    The majority of papers on build optimization were published after 2017, following the release of the popular TravisTorrent dataset. Most studies in the field have been carried out using build datasets, the most widely used of which is TravisTorrent. Finally, the number of publications has been steadily increasing in recent years, demonstrating a growing interest among researchers in this area.

    \item \textbf{\rqtwo}

    Studies on build optimization can be classified into several hierarchical categories grouped into four main topics: 48\% of the studies focused on the \texttt{build duration} and associated costs, 38\% focused on challenges related to the \texttt{build result}, 8\% focused on the general improvement of the \texttt{CI systems} used to execute builds, and 6\% on the \texttt{build data} used for build optimization research.

    \item \textbf{\rqthree}

    Techniques developed to address long build times and their impact can be divided into two categories: (1) build cost reduction techniques, involving the selective execution of certain passing builds, and the acceleration of the build by reusing cached data; and (2) feedback time reduction techniques, which include the prediction of the outcome and duration of the build, and the prioritization of the execution of builds predicted to fail.

    \item \textbf{\rqfour}

    A considerable number of papers have conducted build failure analysis studies. They have found that build failures are associated with code complexity (i.e. large code changes) and mainly result from issues in the code (compilation errors, test failures) and the CI environment. The findings of these studies have been used to develop several techniques for predicting when the build requires maintenance following code changes and for automating build failure repair. 
    
    \item \textbf{\rqfive}

    In addition to build logs, several metrics are used to carry out build studies. We classify these metrics into five categories: code metrics, which are the most commonly used, CI metrics, developer metrics, pull request metrics, and project metrics. They are extracted from various sources such as the VCS and the CI system, and using static code analysis tools. Besides, the calculation of several metrics diverges significantly from one study to another and requires further elucidation to facilitate reuse.

    \item \textbf{\rqsix}

    A total of 8 public build datasets have been published in the literature, which can be leveraged for build failure analysis, build log analysis, build selection techniques, build prediction, and the detection of flaky failures that are unrelated to code. Four of these datasets are specific to the Travis CI system, and four to the Java programming language, highlighting the need for new datasets in a wider range of languages and modern CI tools.
\end{itemize}

\subsection{Replication Package} The list of the identified papers as well as the information noted while analyzing them, are made publicly available in a Google Sheet\footnote{\url{https://tinyurl.com/2npbhsfd}} for future works or replication purposes.

\subsection{Paper Organization} The remainder of this paper is structured as follows: Section \ref{sec:background} presents the background to this study, Section \ref{sec:methodology} presents the methodology followed, and Section \ref{sec:results} presents and discusses the findings. Finally, Section \ref{sec:conclusion} concludes the paper.

\section{Background and Related Work}
\label{sec:background}

The goal of this section is to define and discuss relevant terminologies used in the rest of this paper and present related work that conducted systematic literature reviews on the topic of CI. 

\subsection{Continuous Integration}
\label{sec:background_ci}

Continuous Integration (CI) is the process that automates the execution of different steps required to build and release the software through a pipeline, triggered at the submission of code changes \cite{jin_cost-efficient_2020}. These steps also called jobs, can vary from one development context to another. In general, they include static code analysis, compilation, unit tests, and various tasks that can result in the deployment of a new version into production. So, the CI pipeline aims to ensure the quality of integrated changes, and the fast, reliable creation of a new software version.

For consistency with previous works, we employ the terms \textbf{\texttt{build}} and \texttt{CI} to refer to the \texttt{CI pipeline} in this SLR. 
Indeed, the difference between these terms in the literature remains unclear from one study to another, and they are used interchangeably to refer to the CI pipeline \cite{saidani_bf-detector_2021, lampel_when_2021, gallaba_noise_2018, xia_could_2017, saidani_predicting_2020, esfahani_cloudbuild_2016}. The term \textbf{\texttt{build}} also aligns with the terminology used by popular CI services like Travis CI\footnote{\url{https://www.jenkins.io/doc/book/pipeline}} and Jenkins\footnote{\url{https://docs.travis-ci.com/user/customizing-the-build}}, to refer to the CI pipeline. In contrast, GitHub Actions refers to it as \texttt{workflows}\footnote{\url{https://docs.github.com/en/actions/writing-workflows/quickstart}}, which in reality can be used for tasks unrelated to CI \cite{golzadeh_rise_2022}.

In practice, the build is defined in a configuration file called the \textit{build script} (e.g., \texttt{.travis.yml}) which defines the steps of the pipeline, how they are organized, and tools and commands that need to be run for each step. Developers are encouraged to trigger frequently the build by committing small changes to the VCS \cite{memon_taming_2017, elazhary_uncovering_2022}. When the build execution is completed, i.e. fully executed as opposed to manually canceled, it produces one of two possible results (or outcomes): success or failure. On the one hand, the build is successful when all its sequential jobs are executed successfully. Such build is called a \textit{passing build} \cite{jin_reducing_2021}, and a green color indicates success on each of its jobs. On the other hand, a build failure occurs if any of the build jobs fails. A red color indicates the failed job and the subsequent jobs are not executed as a result of the interruption of the CI pipeline. The build data --- logs and metrics --- associated with the build job executions are used to diagnose the failure to quickly implement fixes toward a successful run of the CI pipeline.

Different dedicated tools known as \textit{CI Systems} are available for setting up and running the build jobs as specified in the build script. These CI systems include Jenkins, Travis CI, CircleCI,  GitLab CI, and GitHub Actions (GHA). GitLab CI and GHA are among the most popular today, both in open source and in industry \cite{widder_im_2018, golzadeh_rise_2022}.
\subsection{Related Work}
\label{sec:related_work}

\begin{table}
\caption{Comparison of this SLR with previous secondary studies}
\label{tab:comparison_slr}
\begin{tabular}{>{\raggedright\arraybackslash}m{4.5cm}  m{2.5cm} >{\raggedright\arraybackslash}m{4.6cm} >{\centering\arraybackslash}m{1.5cm}  >{\raggedleft\arraybackslash}m{1cm}}

\toprule
\textbf{Study} & \textbf{Focus} & \textbf{Findings} & \textbf{\#Papers} & \textbf{Year} \\
\midrule
\citet{stahl_modeling_2014} & CI Practices & Differences in CI practices & 46 &  2014 \\ \midrule
\citet{laukkanen_build_2015} & Build Time & Acceptable and optimal build waiting times & 6 &  2015 \\ \midrule
\citet{shahin_continuous_2017} & CI, CD, CDE & Tools, practices, and challenges & 69 &  2017 \\ \midrule
\citet{laukkanen_problems_2017} & CD & Problems, causes, and solutions & 30 &  2017 \\ \midrule
\citet{pan_test_2021} & Test & Test selection and prioritization techniques using ML & 29 &  2021 \\ \midrule
This study & Build & CI-improving techniques, approaches, and practices & 97 &  2025 \\\bottomrule
\end{tabular}
\end{table}

In this section, we present existing secondary studies that are related to our SLR. Table~\ref{tab:comparison_slr} summarizes the differences in contributions between these existing secondary studies and the present SLR.

\citet{stahl_modeling_2014} studied 46 papers in a systematic literature review to understand and put into evidence the differences in the practices of CI. This SLR highlighted that CI practices cover a large spectrum of variants, making the challenges of one variant potentially inapplicable to another one or the CI practice in general. In addition, the authors proposed a descriptive model that reported the variation points identified in the SLR. 

\citet{laukkanen_build_2015} conducted an interdisciplinary literature review on build waiting time and waiting times in other contexts (e.g., web and computer use), to understand how build duration affects developers. From the analysis of 6 papers, they reported that 10 minutes was an acceptable waiting time for the build, although 2 minutes was considered optimal.

\citet{shahin_continuous_2017} performed a systematic review of the approaches, tools, challenges, and practices of continuous integration, delivery (CD), and deployment (CDE). The study covers DevOps practices in general, and thus reports from the analysis of 69 papers, on the available CI/CD tools, the techniques and challenges related to several different aspects of the development life-cycle, such as code versioning, merge request process, deployment, etc.

\citet{laukkanen_problems_2017} investigated the challenges faced in the adoption of continuous delivery in practice through a systematic literature review of 30 papers. They identified 40 problems classified into seven themes (build design, system design, integration, testing, release, human and organizational, and resource), for which they discovered 28 causes, and synthesized 30 solutions.

\citet{pan_test_2021} focused on test selection and prioritization (TSP) techniques developed using machine learning (ML) algorithms. In this sense, they reviewed 29 papers to summarize the variations in the approaches, used datasets, evaluation metrics, and performances of the existing ML-based TSP techniques. Finally, the authors provided a guideline for classifying future TSP studies.

While the previous secondary studies were conducted on CI practices \cite{stahl_modeling_2014, shahin_continuous_2017, laukkanen_problems_2017}, build time \cite{laukkanen_build_2015}, and testing improvement \cite{pan_test_2021}, they did not focus on the build process, i.e. the improvement of the CI pipeline in general. In this SLR, we focus on the CI build process and identify the approaches, techniques, and practices that exist for optimizing the build. Our SLR covers a larger number of CI build-related papers, with 97 studies published between 2006 and 2024.

\newpage
\section{Methodology}
\label{sec:methodology}

\begin{figure}
  \includegraphics[scale=.7]{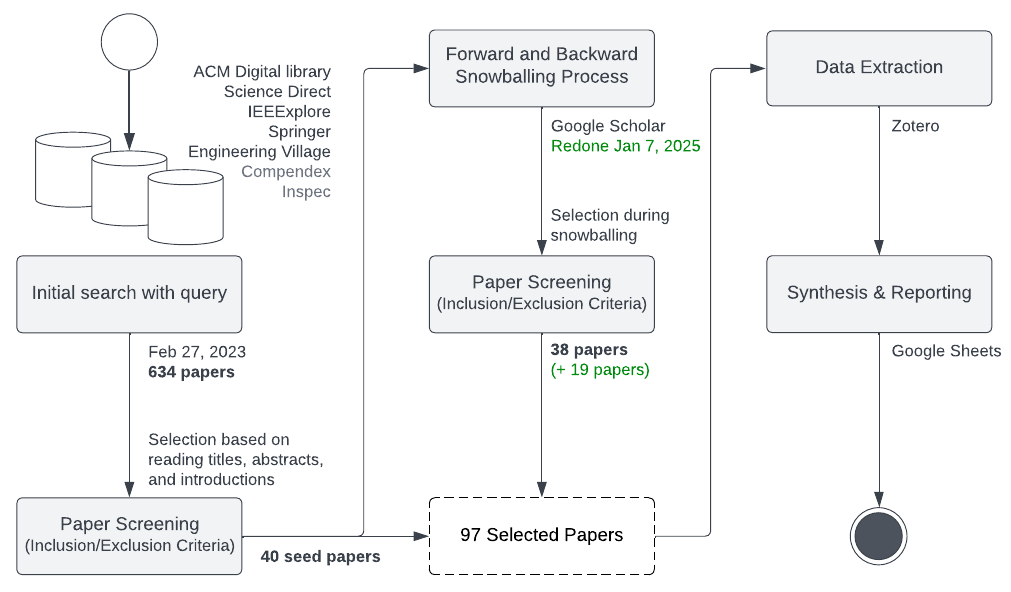}
\caption{Systematic Literature Review Process}
\label{fig:slr_process}
\Description[Steps of the systematic literature review process from initial search with query to synthesis and reporting.]{The systematic literature review process includes initial search, first paper screening, snowballing, second paper screening, then data extraction and synthesis of findings.}
\end{figure}

To conduct our systematic literature review, we follow the approach of ~\citet{kitchenham_guidelines_2007} which is summarized in Figure~\ref{fig:slr_process}. We further discuss the main stages of our methodology in the following subsections. 
 
\subsection{Research Questions}
\label{sec:research_questions}

Since the goal of our SLR is to identify the covered topics by the literature (RQ1-RQ2), the existing solutions (RQ3-RQ4-RQ5), and which datasets are available for research on the area of CI build (RQ6), we design the following research questions:  

\begin{itemize}
    \item[] \rqone
    \item[] \rqtwo 
\end{itemize}

Troubleshooting build failures and excessively long build times are the two main challenges faced by developers in the use of CI systems and that are covered by the literature \cite{hilton_trade-offs_2017}. In order to understand existing practices to address these challenges, we design the following questions:

\begin{itemize}
    \item[] \rqthree
    \item[] \rqfour 
\end{itemize}

ML-based solutions that rely on build data are the most widely used techniques in the literature for improving CI \cite{brandt_logchunks_2020, jin_what_2021}. Hence, we design the next research questions to identify leveraged metrics and datasets available for build optimization research.

\begin{itemize}
    \item[] \rqfive
    \item[] \rqsix 
\end{itemize}

\subsection{Initial Search}
\label{sec:initial_search}

We use string-based search and snowballing techniques to identify the relevant studies in the literature. To formulate the string-based search query, we define search terms that represent the \textbf{population}\footnote{\cite{kitchenham_guidelines_2007} defined the population as (among others) an application area such as an IT system, command, or control system.}: \texttt{Build}, \texttt{Continuous Integration}, and the \textbf{intervention}\footnote{\cite{kitchenham_guidelines_2007} referred to intervention as the software methodology/tool/technology/procedure that address a specific issue.} including popular CI tools names: \texttt{Jenkins, Travis CI, Gitlab CI, GitHub Actions}. In addition, we add as part of the intervention keywords, the \texttt{Build Failure} mechanism, which is designed to provide feedback on CI errors, but also represents one of the major challenges of CI \cite{hilton_trade-offs_2017}. Finally, we complete our search terms with some abbreviations (e.g., \texttt{CI} for \texttt{Continuous Integration}) and synonyms (e.g., \texttt{Breakage} for \texttt{Failure}), to increase the search results.

Furthermore, we have restricted our query results to papers published in a venue whose name contains the keywords \texttt{Software Engineering}, as appearing in the names of international and high-quality software engineering research venues (e.g., ICSE, FSE, and TSE). This is to reduce the noise in our primary results due to a large number of papers (3000+) related to \textit{build} or \textit{construction} in other engineering areas. Obtaining papers published in high-quality venues is also a way for us to ensure the quality and rigor of our analysis. Our finalized \textbf{search query} is as follows.

\begin{center}
    \texttt{(((Build AND (CI OR Continuous Integration) OR ((Jenkins \break OR Travis OR GitLab CI OR GitHub Actions) OR  (Failure OR \break Break OR Broke)))  WN AB) AND ((Software Engineering) WN ST))}
\end{center}

\par As recommended by \citet{kitchenham_guidelines_2007}, we use our search query (initially on the date of Feb. 27, 2023) on six different databases, namely IEEExplore, ACM Digital library, Science Direct, Springer, and Inspec and Compendex provided by the digital meta-library Engineering Village\footnote{\url{https://www.engineeringvillage.com/}} \cite{el-masri_systematic_2020}. The Engineering Village platform offers access to multiple engineering-focused databases (El Compendex, Inspec) that cover papers from major engineering sources and societies such as IEEE library, ASME Digital Collection, Springer, and ACM library \cite{elsevier_engineering_2022}. We include the ACM, Springer, and IEEE databases to ensure that we do not miss any relevant papers even if the Engineering Village indexes these databases. In addition, it is worth noting that Engineering Village provides several features including the \textit{Autostemming}, which allows searching for a root word and its other variations. For example, a search for the word \texttt{Break} will retrieve results also including \texttt{Breakage}.

Also, we opt not to limit the search results to a specific time frame, as relevant studies on build optimization may have been published at any point. This allows us later to capture publication trends in the field of build optimization. Finally, our query resulted in a total of 634 papers. For each paper found, we read the titles and abstracts, then applied the inclusion and exclusion criteria presented in section \ref{sec:criteria}. \textbf{We identify at the end of the first paper screening step 40 seed papers.}

\subsection{Inclusion and Exclusion Criteria}
\label{sec:criteria}

Based on the abstract and title of each paper, we decide to include or exclude it using the following criteria:

\textbf{Inclusion criteria}
\begin{enumerate} 
  \item The paper must be written in English
  \item The paper must be a conference or journal paper
  \item The paper must be a software engineering article
  \item The paper must focus on the CI build process
  \item The study should address at least one way of improving the CI build or present findings for this purpose
\end{enumerate}

\textbf{Exclusion criteria}
\begin{enumerate}
    \item Duplicated papers
    \item The paper is not publicly available
    \item The study is not directly about the CI build, such as studies focusing on testing activities, code compilation process, or just-in-time defect prediction.
    \item The study is related to deployed software (i.e., how to improve software in production)
    \item The study is related to the build but does not focus on its optimization (e.g., introduction of DevOps techniques in software engineering courses \cite{hills_introducing_2020} or general analyses of CI practices)
\end{enumerate}

\subsection{Snowballing Process in Google Scholar}
\label{sec:snowballing}

After identifying the seed papers, we perform the backward and forward snowballing process using Google Scholar\footnote{\url{https://scholar.google.com/}} to capture papers that were not identified using our query. Backward snowballing consists of screening the references list of each included paper, while forward snowballing consists of screening the list of papers that cited each included paper. %
After the snowballing process, we find 38 additional papers (25 from backward snowballing and 13 from forward snowballing). 

Out of the 38 papers, 31 were not identified in our query results due to the criteria on the software engineering venue's name that we have added. Indeed, these 31 papers have been published in other major venues whose names do not contain \texttt{Software Engineering}, such as the International Conference on Mining Software Repositories (MSR), the International Conference on Software Analysis, Evolution and Re-engineering (SANER), and the International Conference on Software Maintenance and Evolution (ICSME).

During the snowballing process, we encountered some special cases that led us to extend our exclusion criteria. Below are the additional exclusion criteria defined:

\begin{enumerate}
  \setcounter{enumi}{5}
  \item Website articles or blogs (e.g. Wikipedia)
  \item Retracted papers and broken links
  \item Books, Doctoral dissertations, Master thesis
\end{enumerate}

\textbf{We identify 78 relevant papers as of February 2023}. After obtaining the final list of papers, we manually examine each of them to answer our research questions.

\subsection{SLR Update using Forward Snowballing in Google Scholar}
\label{sec:slr_update}

In January 2025 we updated our SLR to cover papers published in 2023 and 2024. To do so, we used Forward Snowballing on Google Scholar as recommended by \citet{felizardo_using_2016} and \citet{wohlin_guidelines_2020} for the efficient update of SLRs in software engineering. This forward snowballing is based on the 78 papers identified in February 2023, and the candidate papers are selected using the same inclusion and exclusion criteria. As a result, we find 19 additional relevant studies that we manually examine to include in our initial RQ results.

\textbf{Finally, we obtain a final pool of 97 relevant papers as of January 2025}, that we synthesize in this paper.

\section{Results}
\label{sec:results}

As discussed earlier, the goal of our paper is to understand the state-of-the-art on build optimization. In particular, we address the following research questions: 

\begin{itemize}
    \item[] \rqone
    \item[] \rqtwo
    \item[] \rqthree
    \item[] \rqfour
    \item[] \rqfive
    \item[] \rqsix
\end{itemize}

\subsection{\rqone}

\subsubsection{Motivation.} Similarly to previous SLRs~\cite{malhotra_systematic_2015, el-masri_systematic_2020, gezici_systematic_2022}, we investigate the publication trends over the years in our identified papers, to better understand the interest of the community on the CI build. Such an understanding can be important for researchers who wish to replicate previous studies. For example, prior studies can be replicated on newly available technologies related to the CI pipelines.
Similarly, the trends of build studies might be explained by different factors that we wish to uncover to guide future research work.

\subsubsection{Approach.} To identify the trend of publications, we plot the number of papers published by year and venue type, then analyze the peculiarities (e.g., spikes) in the evolution of the number of publications on build optimization. Furthermore, for each of the selected papers, we identify (1) whether the study relies on build data by checking the presence of \textit{Data Collection}, \textit{Data Extraction}, or \textit{Dataset} section in the paper, and if so (2) the source of the build data that has been used. We define build data source as the build dataset, or if not specified, the open source API\footnote{APIs are used as a standard tool in the technology industry to facilitate data exchange, including build data.} leveraged to collect build data (e.g., GitHub API), or the organization from which build data is collected (e.g., Google, Mozilla). We determine the most relevant data source by examining the information presented in the data section of each paper. Some papers used multiple data sources. In such cases, we count one occurrence for each data source, which results in more data sources than papers in the concerned years. Finally, we analyze using bar plots, the distribution of the papers over the years by build data sources used. For clarity in visualization, we combine data sources used in only one paper into a single category labeled \textit{Others}. This category includes data sources such as Visual Studio Code \cite{rabbani_revisiting_2018}, Apache Software Foundation \cite{maes-bermejo_revisiting_2022}, Ubisoft \cite{olewicki_towards_2022}, StackOverflow \cite{lou_understanding_2020}, and Atlassian \cite{hong_practitioners_2024}. Finally, some studies \cite{gallaba_accelerating_2022, xia_cross-project_2015, cao_forecasting_2017, macho_predicting_2016, kerzazi_why_2014, kawalerowicz_continuous_2023} were evasive about the build data source they used. Hence, we also classify them in the \textit{Others} category.

\begin{figure}
\begin{center}
\includegraphics[scale=0.5]{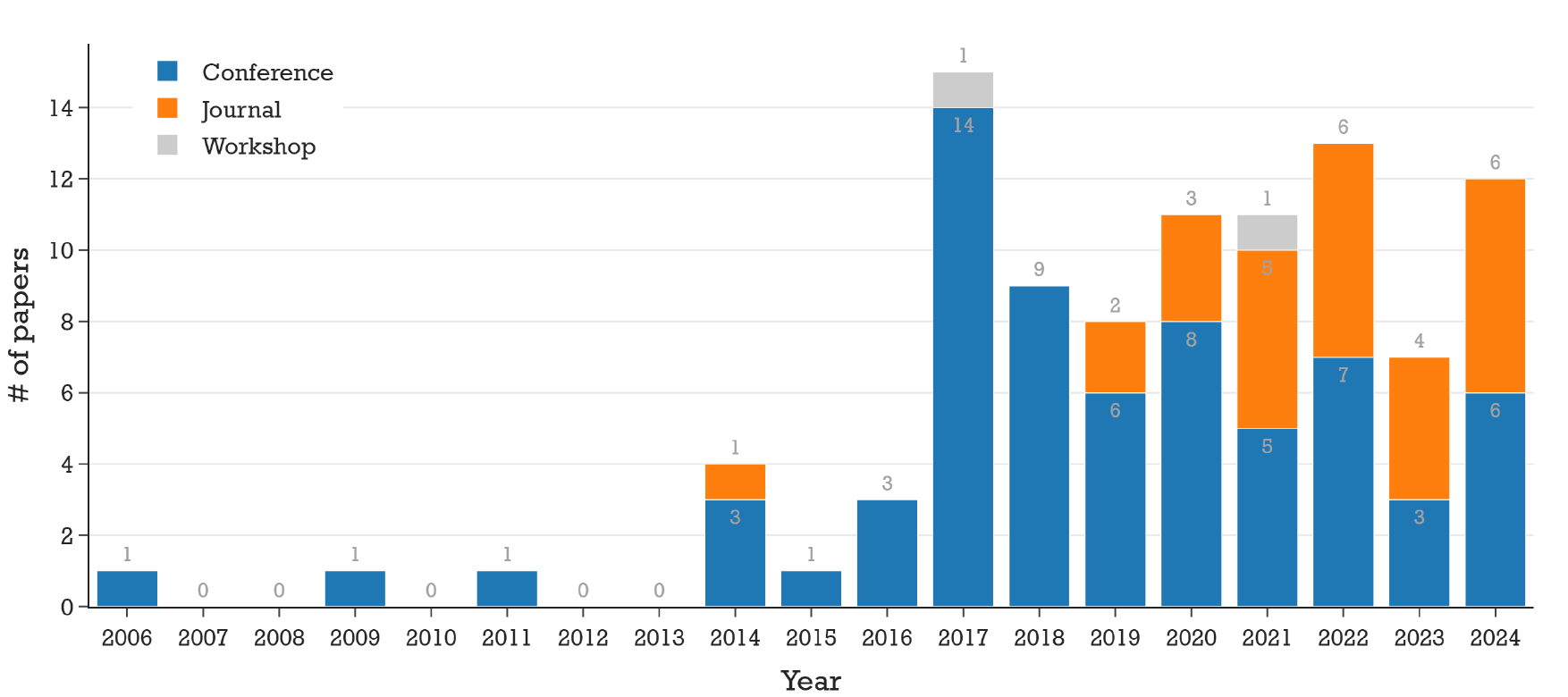}
\end{center}
\caption{Evolution of the number of publications on build optimization by year and type of venue.}
\label{fig:papers_evolution}
\Description[Overall steady increase in the number of publications of build studies]{Overall steady increase in the number of publications of build studies between 2006 and 2024, with a particular spike of 15 publications reached in 2017.}
\end{figure}

\subsubsection{Results.} \textbf{The majority (88.7\%) of the build optimization studies have been published from 2017 onwards, and the publication rate appears to be stable in the past few years}. Figure \ref{fig:papers_evolution} illustrates the yearly publication trends in build optimization research, categorized by type of venue. It shows that build optimization was a very little discussed topic in the literature before 2017, with the number of publications ranging from 0 (in 2007, 2008, 2010, 2012, and 2013) to 4 (in 2014). Out of the 97 studied build papers, only 11 were published before 2017, representing 11.3\% of the papers. Also, no papers on build optimization were identified before 2006, indicating that this topic has gained research interest predominantly within the last decade. Starting in 2017, the number of publications increased significantly, peaking at 15 papers published that particular year, with almost all (14 over 15) published in conference venues. Overall, 86 papers have been published from 2017 to 2024, accounting for 88.7\% of the identified papers. The growing interest in the build topics, which has significantly increased from 2017, is also reflected in consistent annual publications since then. Interestingly, the lowest number of papers published in a single year since 2017 is observed only recently in 2023 with 7 papers. Over the past five years (2020 to 2024), the number of papers published each year has been 11, 11, 13, 7, and 12, respectively, collectively representing over half (55.7\%) of the 97 identified papers. This shows an overall stable publication rate of build studies over the past five years. Before 2019, only a single journal publication was identified. However, from 2019 to 2024, publications are nearly evenly distributed between conference and journal venues, reflecting the growing maturity of research in build optimization. Finally, this SLR was conducted in early 2025, and no build-related papers published in 2025 had been identified. However, we expect the number of publications to evolve in line with the trends observed in recent years.

\begin{figure}
\begin{center}
  \includegraphics[scale=0.5]{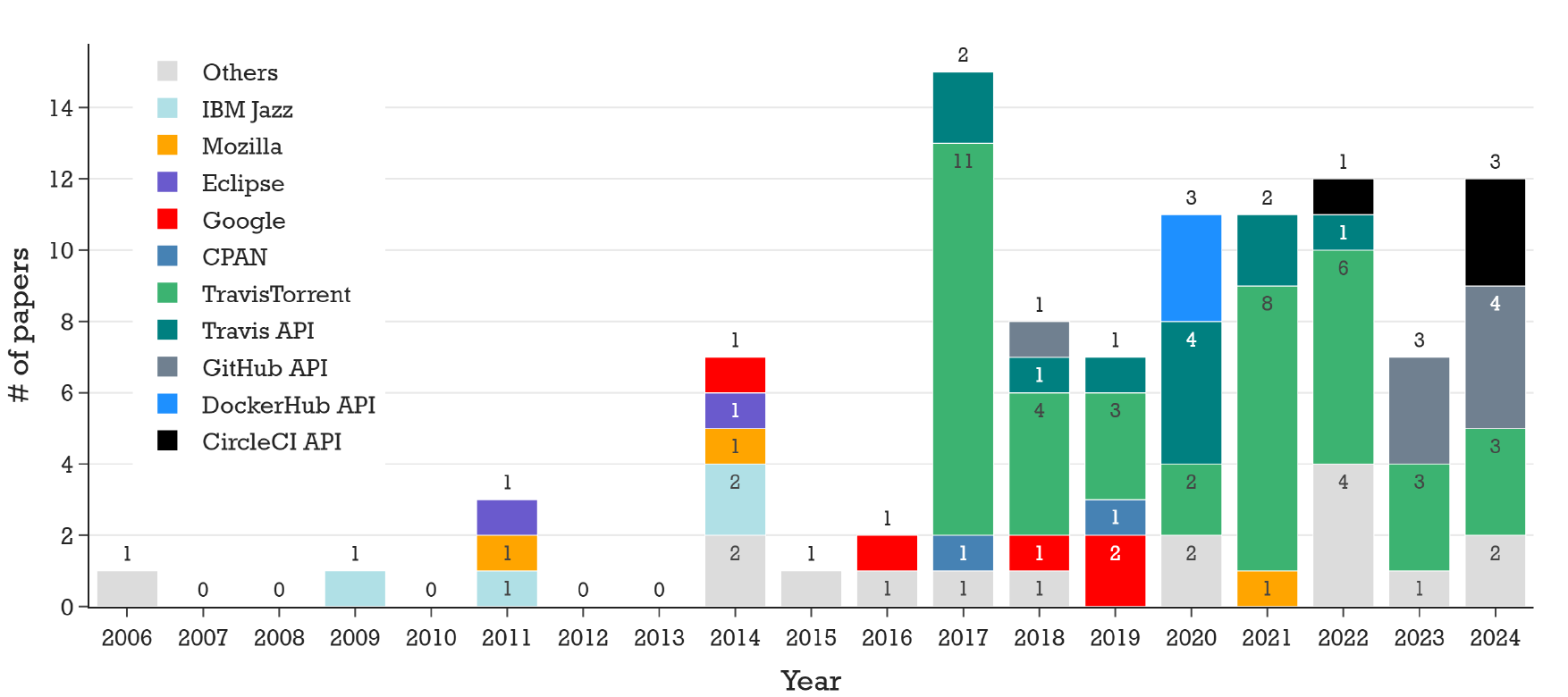}
\end{center}
\caption{Publication frequency over the years and by the build data source used. The annotations on the histogram bars indicate the number of papers using the data source in that particular year.              }
\label{fig:papers_evolution_by_data}
\Description[Overall constant increase in the number of publications with a variety of build data sources]{Overall constant increase in the number of publications with a variety of build data sources including Google, Mozilla, Open Source Software and other datasets, the most important of which is TravisTorrent.}
\end{figure}

\textbf{Existing studies on build optimization are mostly motivated by the open availability of a wealth of historical build data, especially the \textsc{TravisTorrent} dataset which appears to have spearheaded research in the field}. Out of the 97 studied papers, 92 ($\approx$ 95.8\%) relied on historical build data for their studies. Figure \ref{fig:papers_evolution_by_data} presents the distribution over time of build data sources used in these 92 papers. We can observe that the use of build data over time follows approximately the same trend as the number of publications shown in Figure \ref{fig:papers_evolution}. Additionally, Figure \ref{fig:papers_evolution_by_data} reveals that the peak of the number of papers in 2017 is related to the publication in that very year, of \textsc{TravisTorrent}, a large open-source dataset of 2.6 million builds introduced by \citet{beller_travistorrent_2017}. In fact, 11 of the 15 papers (73.3\%) published in 2017 used the \textsc{TravisTorrent} dataset \cite{beller_oops_2017, beller_travistorrent_2017, hassan_change-aware_2017, ni_cost-effective_2017, vassallo_tale_2017, bisong_built_2017, xia_could_2017, luo_what_2017, islam_insights_2017, xia_empirical_2017, dimitropoulos_continuous_2017}. \textsc{TravisTorrent} has also been significantly used in the following years from 2018 to 2024 accounting respectively for 50\%, 42.8\%, 18.2\%, 72.7\%, 46.2\%, 42.8\%, and 25\% of the build data sources used. Besides, out of the 97 papers, a substantial number of papers have also used build data collected from the Travis API (11 papers \cite{saidani_detecting_2021, chen_buildfast_2020, zhang_large-scale_2019, pinto_inadequate_2017, vassallo_-break_2018, vassallo_every_2020, rausch_empirical_2017, barrak_why_2021, zhang_buildsonic_2022, saidani_predicting_2020, brandt_logchunks_2020}), IBM Jazz (4 papers \cite{finlay_data_2014, wolf_predicting_2009, mcintosh_mining_2014, mcintosh_empirical_2011}), and at large organizations such as Google (5 papers \cite{mesbah_deepdelta_2019, seo_programmers_2014, liang_redefining_2018, esfahani_cloudbuild_2016, tufano_towards_2019}) and Mozilla (3 papers \cite{lampel_when_2021, mcintosh_mining_2014, mcintosh_empirical_2011}).

\textbf{Over half (53.6\%) of the build optimization studies have focused on projects using Travis CI as their CI system, although 2024 has seen a noticeable shift towards GitHub Actions and CircleCI}. Figure \ref{fig:papers_evolution_by_data} shows the distribution of the build data sources used in the literature for build optimization studies. It highlights the number of occurrences of each data source for each year. From 2017 to 2024, respectively 86.7\%, 62.5\%, 57.1\%, 54.5\%, 90.9\%, 58.3\%, 42.9\%, and 25\% of the build data sources (per year) are related to the Travis CI system (Travis API and TravisTorrent in shades of green). Overall, 52 papers (53.6\% of the 97 identified papers) used build data collected from projects using Travis CI. This includes 40 papers that specifically employed the \textsc{TravisTorrent} dataset, representing 41.2\% of the 97 identified papers. The other CI systems covered in the literature are Google's CloudBuild (2 papers \cite{esfahani_cloudbuild_2016, tufano_towards_2019}), Microsoft's CloudBuild (1 paper \cite{lebeuf_understanding_2018}), Circle CI (5 papers \cite{gallaba_lessons_2022, gallaba_accelerating_2022, yin_developer-applied_2024, sun_ravenbuild_2024, weeraddana_characterizing_2024, kola-olawuyi_impact_2024}), Mozilla's TaskCluster (1 paper \cite{lampel_when_2021}), Bitbucket Pipelines (2 papers \cite{hong_practitioners_2024, kawalerowicz_continuous_2023}), GitLab CI (1 paper \cite{olewicki_towards_2022}), and GitHub Actions (4 papers \cite{weeraddana_dependency-induced_2024, baral_optimizing_2023, wrobel_using_2023, khatami_catching_2024}). Studies on GitHub Actions (with build data collected from the GitHub API) and CircleCI have mainly emerged over the past two years. In 2024, these studies account for more than half (58.3\%) of the published papers, reflecting a rising interest in these CI systems. These findings suggest that there is potential for further research on newer standard CI systems like GitHub Actions and GitLab CI which are increasingly adopted in both industry and the open-source community \cite{golzadeh_rise_2022}.

\begin{summary}[RQ\textsubscript{1}]
The number of papers on build optimization has significantly increased over the years showing growing interest in this research field. Most build optimization studies leverage historical build datasets, the most widely used in the literature being \textsc{TravisTorrent}. Its release in 2017 explains an unprecedented spike in the number of publications in that year and has enabled many build studies since then. In addition, build optimization research has become increasingly mature, with the numbers of publications in recent years being overall stable and relatively evenly distributed between conferences and journals.
Furthermore, the majority of papers published between  2017 and 2022 focused on Travis CI builds. 
In recent years, particularly in 2024, the focus has shifted toward newer CI systems like GitHub Actions and CircleCI, highlighting opportunities for future studies on builds within these emerging platforms.
\end{summary}

\subsection{\rqtwo}

\begin{figure}
\begin{center}
      \includegraphics[scale=0.4]{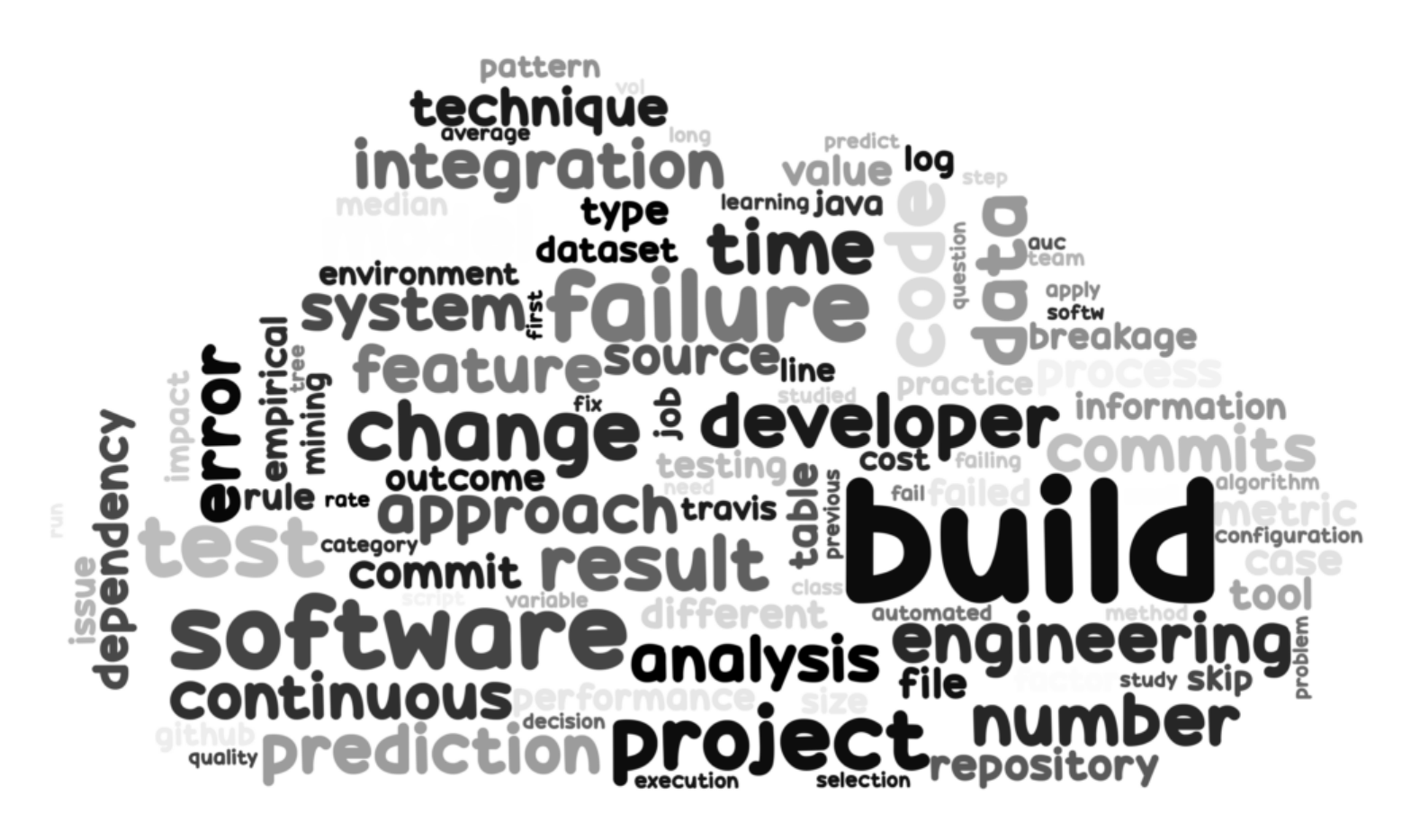}
\end{center}
\caption{Word Cloud generated from the content of the 97 identified papers. The frequent words indicate four primary topics: build quality (\texttt{failure}, \texttt{result}, \texttt{outcome}, \texttt{error}, \texttt{code}, \texttt{test}, \texttt{dependency}, \texttt{issue}), build duration (\texttt{time}, \texttt{cost}, \texttt{prediction}, \texttt{skip}, \texttt{selection}), CI systems (\texttt{system}, \texttt{environment}), and build data (\texttt{data}, \texttt{log}, \texttt{metric}, \texttt{dataset}).}. 
\label{fig:build_studies_wordcloud}
\Description[Word Cloud generated from the content of the 97 identified papers.]{Build studies can be categorized into four primary topics: build quality (\texttt{failure}, \texttt{result}, \texttt{outcome}, \texttt{error}, \texttt{code}, \texttt{test}, \texttt{dependency}, \texttt{issue}), build duration (\texttt{time}, \texttt{cost}, \texttt{prediction}, \texttt{skip}, \texttt{selection}), CI systems (\texttt{system}, \texttt{environment}), and build data (\texttt{data}, \texttt{log}, \texttt{metric}, \texttt{dataset}).}
\end{figure}

\subsubsection{Motivation.} The goal of this research question is to identify the main build-related topics covered by the literature. The findings of this research question will provide insight into the challenges addressed by researchers towards build optimization so practitioners can identify studies that addressed their challenges and researchers can identify topics covered by the literature.

\subsubsection{Approach.} We generate a word cloud (see Figure~\ref{fig:build_studies_wordcloud}) from the content of the 97 papers and manually examine their titles. For this initial analysis, we determine that build optimization studies can be grouped into four primary topics: build duration, build quality, CI systems, and build data. The first three categories are aligned with the main CI build challenges also identified in a previous study~\cite{hilton_trade-offs_2017}, while the last topic covers all papers focusing on build-related data that are used to support studies addressing the three challenges. Next, we review the abstracts of the papers to classify them into the four identified high-level topics. Finally, we conduct a manual examination of the paper content and use the card sorting technique \cite{warfel_card_2004} to further categorize similar papers according to their main goals. Each category was then assigned an appropriate label.

\begin{figure}
\begin{center}
      \includegraphics[scale=.55]{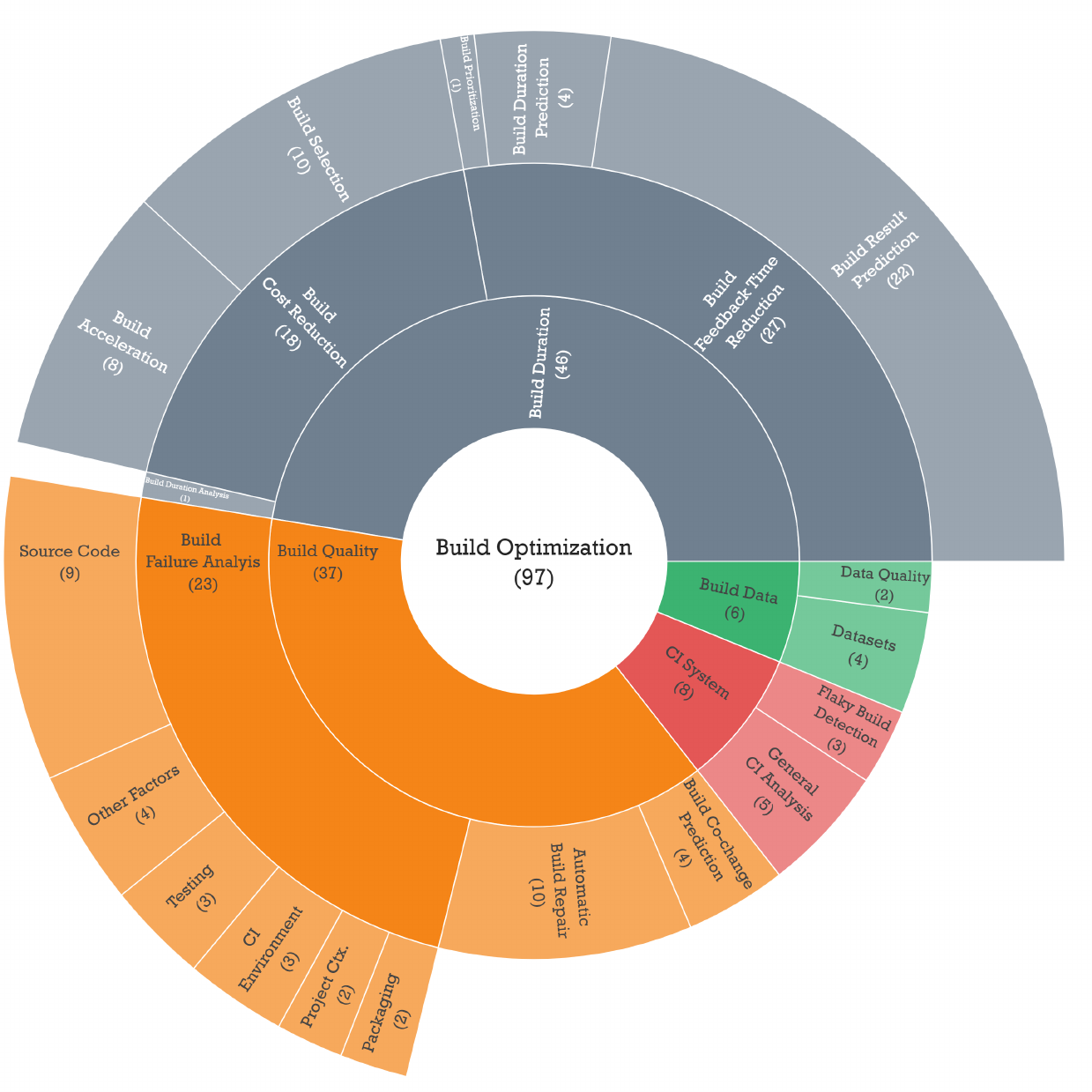}
\end{center}
\caption{Classification of build-related topics illustrated in a Sunburst Diagram. The numbers within the parentheses in the classification nodes represent the count of existing studies on the topic.}
\label{fig:build_categories}
\Description[Build studies classified into four main categories]{Build studies can be classified into four main categories with 35 studies on build duration, 28 on build quality, 8 on CI systems, and 6 on build data leveraged in other studies.}
\end{figure}

\subsubsection{Results.} \textbf{The most frequently discussed topics are related to the build duration, then the build quality}, as shown in Figure~\ref{fig:build_categories}. The figure presents the classification of build optimization studies according to the four identified primary topics. Each primary topic includes multiple sub-topics organized hierarchically. Our technical review of build optimization techniques follows this proposed classification. Subsequent bold statements provide detailed discussions on each primary topic, offering critical information used to address the additional research questions outlined in Section \ref{sec:research_questions}.

\textbf{On the topic of build duration, the literature focuses on reducing the cost of running long builds, and investigates techniques for obtaining build results more quickly to minimize waiting times.} The most important topic of CI build addressed in the literature is the \texttt{build duration}, accounting for 47.4\% of the studies. As shown in Figure \ref{fig:build_categories}, two main goals have been identified in this research direction. The first goal --- \texttt{build cost reduction} --- is to reduce the high costs of frequently running long builds (estimated in millions of dollars at Google and Microsoft \cite{zhang_buildsonic_2022}). To do so, prior studies \cite{jin_which_2022, jin_reducing_2021, jin_cost-efficient_2020, saidani_detecting_2021, abdalkareem_which_2021, abdalkareem_machine_2021, beheshtian_software_2022, weeraddana_dependency-induced_2024, jin_span_2023, kamath_combining_2024} investigated \texttt{build selection} to help practitioners identify the right builds to execute and skip the others to save costs. The right builds are the ones most likely to fail and reveal issues in the software version. Other studies \cite{esfahani_cloudbuild_2016, celik_build_2016, zhang_buildsonic_2022, gallaba_accelerating_2022, randrianaina_towards_2022, yin_developer-applied_2024, khatami_catching_2024, baral_optimizing_2023} focused on \texttt{build acceleration} by reusing cached data from previous executions or repairing CI performance smells. 
The second goal of existing research on build duration is to minimize the feedback waiting time --- \texttt{build feedback time reduction} --- for increased productivity. To do so, prior studies such as\footnote{A comprehensive list of build result prediction studies is provided in RQ\textsubscript{3}} \cite{chen_buildfast_2020, santolucito_learning_2022, wu_using_2020, saidani_improving_2022, kawalerowicz_continuous_2023, wang_commit_2024, sun_ravenbuild_2024} developed techniques for \texttt{build result prediction} before its execution, so that depending on the predicted result, practitioners can either decide to skip the execution or investigate their code for potential errors. Another line of work \cite{cao_forecasting_2017, bisong_built_2017, tufano_towards_2019, ghaleb_empirical_2019} explored \texttt{build duration prediction} to predict the total build time, enabling practitioners to better manage context switching from one task to another. Finally, a prior study \cite{liang_redefining_2018}  investigated \texttt{build prioritization} consisting of prioritizing the execution of builds prone to failure, so that practitioners could receive early feedback from the CI. Another study \cite{weeraddana_characterizing_2024} conducted a \texttt{build duration analysis} of timeout builds to understand the associated factors. RQ\textsubscript{3} discusses in more detail the techniques used for optimizing build duration.

\textbf{On the topic of build quality, existing studies have sought to understand and address build failures in order to improve software quality.} The second most discussed topic in the literature is the \texttt{build quality}, addressed in 37 papers accounting for 38.1\% of the studies. On the one hand, a substantial number of studies (23 papers) focused on \texttt{build failure analysis} to understand the circumstances under which builds fail. In this sense, prior studies investigated the root cause of pipeline failures at different levels, including, for example\footnote{The build failure factors are discussed in detail in RQ\textsubscript{4}}, \texttt{source code} \cite{barrak_why_2021, seo_programmers_2014, zhang_large-scale_2019},\texttt{testing} \cite{rausch_empirical_2017, beller_oops_2017, vassallo_tale_2017} and \texttt{packaging} \cite{wu_empirical_2020, henkel_learning_2020}. On the other hand, other studies have developed several techniques to improve the quality of the build by mitigating build failures. For this purpose, a line of work \cite{mcintosh_empirical_2011, macho_predicting_2016, mcintosh_mining_2014, xia_cross-project_2015} investigated \texttt{build co-change prediction} to automatically detect when the build script requires maintenance following code changes, assisting practitioners in the build maintenance. Other studies \cite{macho_automatically_2018, vassallo_-break_2018, hassan_hirebuild_2018, lou_history-driven_2019, mesbah_deepdelta_2019, hassan_tackling_2019, vassallo_every_2020, lou_understanding_2020} have proposed \texttt{automatic build repair} techniques that promptly correct code-related issues (e.g., syntax errors, dependency issues) to remediate build failures. RQ\textsubscript{4} discusses in more detail the techniques proposed to address build failures.

\textbf{Other studies (8.2\%) in the literature have focused on the topic of CI systems, to improve the reliability and overall performance of build executions.} The third primary topic addressed in the literature is \texttt{CI System} and includes 8 (8.2\%) of the identified studies. This topic encompasses, on one hand, \texttt{general CI analysis} of issues related to the usage of CI platforms used to run the builds. For instance, \citet{pinto_inadequate_2017} highlighted that time pressure and inadequate testing are two main factors of build failures in organizations, where CI gives a false sense of confidence. Then, further studies \cite{vassallo_enabling_2019, gallaba_lessons_2022} investigated the problems in the CI system that cause irregular build failures (i.e. flaky builds) and found a wide variety of reasons, including user cancellation, infrastructure availability problems, configuration errors, and timeouts. Other studies \cite{lebeuf_understanding_2018, jin_what_2021} have evaluated existing techniques for optimizing CI regarding various aspects such as execution costs, failure observation, and feedback time. On the other hand, prior studies \cite{lampel_when_2021, olewicki_towards_2022, moriconi_automated_2022} explored \texttt{flaky build detection} approaches for identifying and diagnosing irregular build failures (also known as flaky builds), to improve the reliability of build execution. The techniques used are also further discussed in RQ\textsubscript{4}.

\textbf{A small proportion (6.2\%) of the papers focused on providing datasets or analyzing the quality of data used in build optimization research.} The fourth primary topic investigated in the literature --- \texttt{build data} --- encompasses studies that introduced build datasets or discussed the quality of build data. In this sense, prior studies \cite{beller_travistorrent_2017, brandt_logchunks_2020, sulir_large-scale_2020, jin_cibench_2021} presented large and publicly available build datasets that can be leveraged for different build optimization purposes. Other studies \cite{gallaba_noise_2018, ghaleb_studying_2021} discussed the quality of build data used, and the potential bias that could result from using noisy datasets. Their findings are further discussed in RQ\textsubscript{5}.

\begin{summary}[RQ\textsubscript{2}]
The overwhelming majority (85.6\%) of the existing studies have focused on minimizing build duration and tackling build failures. In particular, studies on build duration can be classified into topics such as build selection, build acceleration, build prediction, and build prioritization; while studies on build failures can be classified as build failure analysis, build maintenance prediction, and automatic build repair. In addition, a limited number of papers (8.2\%) have focused on improving the reliability and performance of CI systems. Finally, a few studies (6.2\%) have focused on the build data used for research in this field.
\end{summary}

\subsection{\rqthree}

\subsubsection{Motivation.} Since we observe in the previous research question that build duration is among the most studied build-related topics, the objective of this research question is to dig deeper into the techniques investigated in the literature so practitioners can identify solutions to their problems related to build duration.

\subsubsection{Approach.} To identify the techniques that are used to address the problem related to the build duration, we manually analyze the studies on build duration by looking at the approach of each of the relevant papers. Figure \ref{fig:build_categories} shows that the studies on build duration are distributed according to two main goals. The first line of work sought to reduce the huge infrastructure and energy costs involved in running long builds. Meanwhile, the second set out to explore solutions aimed at obtaining faster feedback in order to maintain team productivity. We extract and synthesize the techniques that have been developed in line with these two main goals.

\subsubsection{Results.} \textbf{Several techniques, mostly based on machine learning prediction algorithms, have been developed to reduce the cost and the feedback time of long builds}, as summarized in Table \ref{tab:build_duration_techniques}. 
As indicated in the table, 54.3\%, 10.9\%, 8.7\%, and 8.7\% of the studies focusing on build duration respectively used machine learning (ML), search algorithms, information retrieval, and caching strategies, highlighting these as the four most commonly leveraged techniques in the literature to address build duration. Other studies used deep learning (6.5\%), statistical analysis (6.5\%), and rules-based (4.3\%) techniques. We further discuss the techniques in the following bold statements.

\begin{footnotesize}
\begin{longtable}{p{0.17\linewidth} m{0.16\linewidth} p{0.53\linewidth} p{0.05\linewidth}}
\caption{Techniques investigated in the literature to address long build duration.}
\label{tab:build_duration_techniques}\\
    \toprule
        \textbf{Topic} & \textbf{Technique} & \textbf{Description} & \textbf{Ref.} \\
    \midrule
 \endfirsthead

 \multicolumn{4}{c}{Continuation of Table \ref{tab:build_duration_techniques}}\\
    \toprule
        \textbf{Topic} & \textbf{Technique} & \textbf{Description} & \textbf{Ref.}\\
    \midrule
 \endhead

    \bottomrule
 \endfoot

    \bottomrule
 \endlastfoot

\multirow{16}{\linewidth}{Build Acceleration (8)} & \multirow{8}{\linewidth}{Information Retrieval (4)} & \textit{BuildSonic}. Developed a tool to detect and repair performance-related build script smells using a created catalog of 20 performance smells. & \cite{zhang_buildsonic_2022} \\ \cmidrule{3-4}

& & \textit{OptCD}. Identified build tasks generating unnecessary content (e.g., unused files) and proposed an updated build config. file version excluding them.  & \cite{baral_optimizing_2023} \\\cmidrule{3-4}

& & Developed a toolset to detect GitHub Actions workflow configuration smells using a created catalog including 10 performance-related smells.  & \cite{khatami_catching_2024} \\\cmidrule{3-4}

& & Studied developer-applied CI accelerations and identified 16 patterns.  & \cite{yin_developer-applied_2024} \\
        
        \cmidrule{2-4}
 
 & 
        \multirow{7}{\linewidth}{Cache Strategy (4)} & Cached dependencies files to lazily reuse them for next build executions. & \cite{celik_build_2016} \\ \cmidrule{3-4}
        & & Designed and built a distributed cache service to reuse dependencies from previous builds. & \cite{esfahani_cloudbuild_2016} \\ \cmidrule{3-4}
        & & Cached dependency graph and CI artifacts to skip steps in the next builds. & \cite{gallaba_accelerating_2022} \\ \cmidrule{3-4}
        & & Investigated the approach of incremental build using previous builds on different software configurations. & \cite{randrianaina_towards_2022} \\ 

        \midrule

\multirow{18}{\linewidth}{Build Selection (10)} & \multirow{4}{\linewidth}{Rules Based (2)} & Skipped commits based on defined rules on changes that do not necessitate running a build such as comments or formatting changes. & \cite{abdalkareem_which_2021} \\ \cmidrule{3-4}
                & & \textit{Dep-sCImitar}. Skipped builds triggered by commits of updates to unused dependencies having a history of successful builds. & \cite{weeraddana_dependency-induced_2024} \\ \cmidrule{2-4}

                & \multirow{12}{\linewidth}{Machine Learning (7)}  & \textit{HybridBuildSkip}, \textit{SmartBuildSkip}, \textit{PreciseBuildSkip}. Predicted first failures and ran next builds until a passing build to minimize escaped failures. & \cite{jin_cost-efficient_2020, jin_reducing_2021, jin_which_2022} \\ \cmidrule{3-4}
                & & Predicted whether a commit can be skipped using 23 commits level features. & \cite{abdalkareem_machine_2021} \\ \cmidrule{3-4}
                & & Grouped the execution of builds according to a predicted risk threshold, and skipped the safe builds. & \cite{beheshtian_software_2022} \\ \cmidrule{3-4}
                & & \textit{HybridCISave}. Combined \textit{HybridBuildSkip} which leverages existing techniques as features to skip builds predicted to pass, with \textit{HybridTestSkip} executed otherwise to skip tests predicted to pass.  & \cite{jin_span_2023} \\ \cmidrule{3-4}
                & & Introduced a hybrid approach involving the prediction of builds to skip and grouping build commits for executions. & \cite{kamath_combining_2024} \\ \cmidrule{2-4}
                
                & \multirow{2}{\linewidth}{Search Algorithm (1)} & Developed commit skipping rules by searching for the best features and threshold values combination. & \cite{saidani_detecting_2021} \\
               
                \midrule
        
        \multirow{2}{\linewidth}{Build Prioritization (1)} & 
                \multirow{2}{\linewidth}{Statistical Analysis (1)} & Prioritized the execution of commits that are likely to fail based on statistics about historical commit execution data. & \cite{liang_redefining_2018} \\

                        \midrule
        
        \multirow{7}{\linewidth}{Build Duration Prediction (4)} & 
                \multirow{2}{\linewidth}{Machine Learning (2)} & Predicted build job duration using historical build data and ML algorithms. & \cite{bisong_built_2017} \\ \cmidrule{3-4}
                & & Investigated the causes of long builds using statistical and predictive models. & \cite{ghaleb_empirical_2019} \\ \cmidrule{2-4}
                
                & \multirow{3}{\linewidth}{Statistical Analysis (2)} & Analyzed build dependency graphs (BDG) to forecast build job duration. & \cite{cao_forecasting_2017} \\ \cmidrule{3-4}
                & & Estimated build duration by analyzing historical execution time statistics and the impact of changes on the BDG. & \cite{tufano_towards_2019} \\ 
                
                \midrule

 &  & Predicted build results using historical data about social communications within teams. & \cite{hassan_using_2006, wolf_predicting_2009} \\ \cmidrule{3-4}
        & & Predicted build results using a data stream of historical code change metrics. & \cite{finlay_data_2014} \\ \cmidrule{3-4}
     \multirow{34}{\linewidth}{Build Result \break Prediction (22)}   & \multirow{20}{\linewidth}{Machine Learning (15)} & Predicted build results using features related to source code changes. & \cite{hassan_change-aware_2017} \\ \cmidrule{3-4}
        & & Trained cascaded classifiers to deal with build data imbalance issues. & \cite{ni_cost-effective_2017} \\ \cmidrule{3-4}
        & & Investigated different data selection strategies to improve the performance of predictive models trained using cross-project historical data. & \cite{xia_could_2017, xia_empirical_2017} \\ \cmidrule{3-4}
        & & Used a set of predictive models that learn from build events as they occur over time to limit the effects of concept drift. & \cite{ni_acona_2018} \\ \cmidrule{3-4}
        & & Developed a semi-supervised AUC optimization technique that learns from a build data stream to predict build outcomes. & \cite{xie_cutting_2018} \\ \cmidrule{3-4}
        & & Parsed Dockerfiles into tokens which are transformed into numerical features used to train container build failure prediction models. & \cite{wu_using_2020} \\ \cmidrule{3-4}
        & & \textit{BuildFast}. Predicted build outcome using failure-specific features of historical build activity. & \cite{chen_buildfast_2020} \\ \cmidrule{3-4}
        & & Predicted build results using textual features from CI configuration files, to locate the root cause of errors. & \cite{santolucito_learning_2022} \\ \cmidrule{3-4}
        & & Extracted textual features of commits to predict build failures and locate the root cause of errors. & \cite{al-sabbagh_predicting_2022} \\ \cmidrule{3-4}
        & & Developed context- and dependency-aware features to predict build failures, achieving performance surpassing BuildFast. & \cite{sun_ravenbuild_2024} \\ \cmidrule{3-4}
        & & \textit{CBOP}. Developed a toolset to continuously predict build results on code submission and provide developers with timely feedback in the IDE. & \cite{kawalerowicz_continuous_2023} \\ 
        
        \cmidrule{2-4}

        & \multirow{3}{\linewidth}{Search Algorithm (4)} & Classified build results using rules learned from historical code smells data. & \cite{saidani_toward_2021} \\ \cmidrule{3-4}
        & & \textit{BF Detector}, \textit{MOGP}. Developed prediction rules by combining key CI metrics and threshold values identified using search-based algorithms like NSGA-II. & \cite{saidani_prediction_2020, saidani_predicting_2020, saidani_bf-detector_2021} \\ 
        \cmidrule{2-4}

        & \multirow{8}{\linewidth}{Deep Learning \& LLMs (3)} & \textit{DL-CIBuild}. Improved build prediction with an LSTM model, and optimized its hyper-parameters using genetic algorithms. & \cite{saidani_improving_2022} \\ \cmidrule{3-4}
        & & \textit{GitSense}. Combined several commit artifact data into textual features to train a transformer-based model that outperformed existing techniques: \textit{MOGP} \cite{saidani_predicting_2020}, \textit{BuildFast} \cite{chen_buildfast_2020}, SmartBuildSkip \cite{jin_cost-efficient_2020}, and HybridCISave \cite{jin_span_2023} & \cite{wang_commit_2024} \\ \cmidrule{3-4}
        & & Developed a hybrid approach using LSTM-GRU models to improve build predictions; and dynamic build success optimization based on 4 key metrics. & \cite{benjamin_enhancing_2024} \\

        \midrule
        
        \multirow{2}{\linewidth}{Build Duration Analysis (1)} & \multirow{2}{\linewidth}{Machine Learning (1)} & Analyzed timeout builds to identify key factors, with build history and timeout tendency emerging as the most significant. & \cite{weeraddana_characterizing_2024} \\
        
\end{longtable}
\end{footnotesize}

\textbf{Different techniques, from rules-based to machine-learning techniques, were evaluated for skipping builds.} Certain types of changes made to the repository files (e.g., documentation updates) unnecessarily trigger build execution. To address this, \citet{abdalkareem_which_2021} defined rules to decide whether a change requires running the build or can be skipped, according to the type of committed changes (e.g., skip builds for changes to code comments, documentation files, code formatting, meta files, and release preparation code). \citet{abdalkareem_machine_2021} improved their rules-based solution with a Decision Tree ML model that used 23 commit-related features to predict whether a commit can be skipped. Later, \citet{weeraddana_dependency-induced_2024} developed \textit{Dep-sCImitar}\footnote{\url{https://www.npmjs.com/package/dep-scimitar}}, a rules-based \texttt{npm} package to skip builds triggered by unused dependency updates with a history of successful builds, saving 68.34\% of build time in such cases.

Skipping builds can come up with a risk of missing relevant builds, i.e. builds that would reveal a failure. To overcome such an issue, other studies \cite{jin_cost-efficient_2020, jin_reducing_2021} leveraged an ML model to detect first failures (i.e. failures not preceded by another one), in which case they stop using the model and run the subsequent builds until a successful build is observed. At this point, they resume running the model while skipping some of the builds predicted as passing to save costs. This way, they have improved the safety of their build selection process, by maximizing failure observations. For the same build selection purpose, another line of work -- \citet{beheshtian_software_2022, kamath_combining_2024} -- explored the combination of techniques for (1) grouping the build commits to be executed and (2) predicting the builds to be skipped. Further, \citet{jin_which_2022} extended the commit skip rules of \citet{abdalkareem_which_2021} with new rules and used 
ML for build selection. Such a model uses the rules themselves as features. Their approach outperformed the existing rules-based \cite{abdalkareem_which_2021} and ML-based techniques \cite{abdalkareem_machine_2021, jin_cost-efficient_2020}. \citet{saidani_detecting_2021} also explored methods for automatically generating rules to identify CI builds that can be skipped, using a search-based algorithm known as the Strength-Pareto Evolutionary Algorithm (SPEA-2). However, this latter technique was found to be ineffective in 52\% of the studied projects by \citet{jin_span_2023}, as it took more time to generate skipping decisions than to execute the build itself.

Ultimately, \citet{jin_span_2023} integrated several existing build selection techniques \cite{abdalkareem_which_2021, abdalkareem_machine_2021, jin_cost-efficient_2020} along with test selection techniques as features to propose \textit{HybridBuildSkip}, which outperformed state-of-the-art (SOTA) approaches. They further developed \textit{HybridBuildSave}, which combines \textit{HybridBuildSkip} for skipping builds predicted to pass and \textit{HybridTestSkip} for skipping tests predicted to pass (when a build failure is predicted), maximizing build cost savings while maintaining satisfactory recall (i.e. execution) of relevant builds.

\textbf{Several caching strategies and performance smells repair techniques based on information retrieval have been investigated to accelerate builds instead of skipping them.} \citet{celik_build_2016} designed a build execution approach that caches dependencies files during first build executions, and later lazily identifies and retrieves necessary cached files. As a result, they retrieved dependencies 93.81\% faster than the \textit{Maven} build tool. \citet{esfahani_cloudbuild_2016} leveraged a similar cache strategy to accelerate builds in the context of large-scale distributed systems at Microsoft. They developed an in-house build tool including a distributed cache service to execute build jobs only when necessary, resulting in up to 10 times faster build duration. \citet{zhang_buildsonic_2022} focused on fixing performance-related CI configuration smells (e.g., caching not enabled) to speed up the build process by 12.4\%. \citet{khatami_catching_2024} also developed a toolset to detect GHA workflow configuration smells (including 10 performance-related smells), achieving promising results (F1-score: 0.81--1). To support future automated performance smells repair, \citet{yin_developer-applied_2024} studied developer-applied CI accelerations in CircleCI and found 16 acceleration patterns for reuse. On another approach, \citet{baral_optimizing_2023} developed \textit{OptCD} to detect unnecessarily generated files and their associated GitHub Actions build steps, then used ChatGPT to suggest patches to remove these steps from the build script, speeding build time by 7\%.

Directly reusing cached dependency files is constrained by programming language and build tool peculiarities. Furthermore, this technique is less effective when using container technologies (e.g., Docker), with which build environments are frequently recreated to restart with a clean one. To fill this gap, \citet{gallaba_accelerating_2022} cached the container build environment during the execution of a first build and created the next containers on top of the cached container to avoid redundant steps of container environment configuration. This technique reduced by half the duration of 74\% of the studied builds.
\citet{randrianaina_towards_2022} explored the approach of incremental build for software with large build configuration spaces. They sought to find the right order to execute the build for a given configuration space, by reusing commonalities from previous builds on the next different configurations. The approach reduced up to 66\% of build time. 

Despite the availability of these techniques, a recent study by \citet{yin_developer-applied_2024} on developer-applied CI accelerations (in GitHub projects using CircleCI) reveals that, in practice, developers still primarily rely on checking specific file changes (e.g., documentation) and CI environment variables to skip certain steps for build acceleration.

\textbf{Various techniques from prioritizing commits execution to sophisticated prediction models such as LLMs have been evaluated to predict build results.} \citet{liang_redefining_2018} used a queue prioritization technique to prioritize the execution of the commits that are most likely to fail. Such commits are identified based on statistical analysis of historical data. In general, the literature mainly leveraged ML classifiers trained on different build data to predict build results. Early studies \cite{hassan_using_2006, wolf_predicting_2009} used teams' social communications features to train ML classifiers. To improve build prediction models, prior studies \cite{finlay_data_2014, hassan_change-aware_2017, saidani_toward_2021} investigated code change metrics as the main prediction features. \citet{chen_buildfast_2020} have even designed historical and failure-specific features to improve the performance of their predictive model, outperforming existing SOTA build prediction techniques \cite{hassan_change-aware_2017, ni_cost-effective_2017, xie_cutting_2018}. \citet{wu_using_2020} investigated the prediction of container build failures using semantic features of Dockerfiles to train ML classifiers. Finally, other studies \cite{santolucito_learning_2022, al-sabbagh_predicting_2022} trained tree-based classifiers on specific code changes metrics (e.g., CI configuration smells features, textual features of file changes) to enable the localization of the root cause of failures following a failure prediction.

Predicting the outcome of a build requires a significant amount of historical data, which is not usually available on new projects. To address this issue, \citet{xia_could_2017} studied the performance of ML classifiers on build data collected across 126 open-source projects, and argued that build prediction models yield acceptable results on projects on which they have not been trained. \citet{xia_empirical_2017} improved the prediction performance by 45\% using the \textit{Bellwether strategy} \cite{chen_bellwether_2006} for project-level data selection. This data selection strategy consists of finding the subsets of data least costly to train (called "bellwethers") and sufficient to maximize prediction accuracy. 

Prior studies overlooked the temporal dimension of build data, which could enhance predictive accuracy. To fill this gap, \citet{saidani_improving_2022} investigated the performance of the deep learning LSTM model %
for cross-project build prediction and used a genetic algorithm to optimize its hyper-parameters. The model outperformed the SOTA ML models both in online scenarios and cross-project prediction on the TravisTorrent dataset. \citet{benjamin_enhancing_2024} improved build outcome prediction techniques that consider the temporal nature of build data, using a hybrid approach based on LSTM and Gated Recurrent Unit (GRU) models. They also identified four interdependent key metrics ($B_{10}$, $B_{6}$, $B_{13}$, and $C_{9}$ in Tables~\ref{tab:ci_metrics} and ~\ref{tab:code_metrics}), for which they suggested continuous monitoring and real-time threshold-based actions to achieve dynamic build success optimization.

Several challenges, such as data imbalance and concept drift, hinder the achievement and maintenance over time of high performance in build outcomes prediction. Focusing on the build data imbalance issue, a line of work \cite{saidani_prediction_2020, saidani_predicting_2020} used the Non-dominated Sorting Genetic Algorithm (NSGA-II) \cite{deb_fast_2002} to differentiate effectively between the minority and the majority classes, respectively \texttt{failure} and \texttt{success}. To meet the concept drift challenges, \citet{finlay_data_2014} used a concept drift detection mechanism -- the adaptive sliding window \cite{bifet_learning_2007} -- to learn from a simulated stream of build data, and incrementally update the weights of their Hoeffding Tree (HT) classifier \cite{hulten_mining_2001}. \citet{xie_cutting_2018} improved this technique using a semi-supervised online AUC optimization algorithm which also resulted in faster model training. Further, \citet{ni_acona_2018} developed an active learning approach including a pool of weighted ML classifiers. The internal and external weights of the models are updated using a real-time stream of build data to maximize prediction performance over time. In the same direction, \citet{kawalerowicz_continuous_2023} developed a toolset for continuous build outcome prediction (CBOP) --- leveraging tree-based ML models --- that provides real-time feedback to developers in their VS Code workspace. Interestingly, they found that developers tend to cause more build failures when using CBOP compared to not using it, highlighting practitioners' concerns about overconfidence and prediction accuracy \cite{kawalerowicz_continuous_2023, hong_practitioners_2024}.

In some scenarios, traditional code and build-related features alone are insufficient to capture the factors influencing build outcomes, leading to poor prediction performance. This issue is particularly evident in software games, where non-code assets such as animation, sound, and graphical scene files are modified more frequently than code and provide better explanations for failures \cite{sun_ravenbuild_2024}. To overcome this limitation, \citet{sun_ravenbuild_2024} developed context-, relevance-, and dependency-aware features, which incorporated, resulted in better performance than \textit{BuildFast} \cite{chen_buildfast_2020} in predicting build outcomes across 22 open-source projects. \citet{wang_commit_2024} combined various commit artifact data --- such as commit message, code changes, and name of changed files --- into numerical textual features, from which the best feature representations (known as \textit{embeddings}) are learned using a transformer-based encoder. The outputs of two multi-layer perceptrons trained on the embeddings are then combined to predict build results. This approach named \textit{GitSense} outperformed several SOTA techniques including \textit{MOGP} \cite{saidani_predicting_2020}, \textit{BuildFast} \cite{chen_buildfast_2020}, \textit{SmartBuildSkip} \cite{jin_cost-efficient_2020}, and \textit{HybridCISave} \cite{jin_span_2023}.

\textbf{The literature leveraged the analysis of build dependency graphs and machine learning techniques to predict the build duration.} \citet{cao_forecasting_2017} investigated leveraging the build dependency graphs and statistical analysis to estimate the duration of incremental build jobs. The technique accurately predicted job duration at an error of 10 seconds, for 94\% of the commits that did not change the build dependency graph. \citet{tufano_towards_2019} proposed a similar approach to estimate build time in a distributed and caching build system (Microsoft's CloudBuild~\cite{esfahani_cloudbuild_2016}), based on determining the longest critical path in the dependency graph.

\citet{bisong_built_2017} compared several ML models trained on TravisTorrent for the prediction of build duration, so that developers can better manage their schedule and avoid wasted waiting time. The Random Forest (RF) and the Cubist \cite{kuhn_cubist_2012} models delivered the best prediction performances. \citet{ghaleb_empirical_2019} also used ML techniques (mixed-effects logistic regression) to investigate the main factors explaining long build times in 67 projects. Their study revealed that long build times are associated not only with high code and test density, but also with various misuses of CI configurations, such as systematic updates of dependencies even when they are not needed, repeated retries of failed commands, or not enabling caching when possible.

\begin{summary}[RQ\textsubscript{3}]
To address the challenge of overly long builds, several techniques mostly based on machine learning algorithms have been investigated in the literature. These techniques can be classified into two main categories, each addressing the two major issues caused by such builds: (1) cost reduction techniques including (a) partial, selective, or grouped execution of builds predicted to pass, and (b) build acceleration using caching strategies and fixing performance smells in build scripts; (2) feedback time reduction techniques including (a) prioritizing failing commits and (b) predicting build outcomes and (c) durations. We recommend further studies on the factors influencing build duration and the development of better techniques for prioritizing build executions and repairing build script performance smells.
\end{summary}

\subsection{\rqfour}

\subsubsection{Motivation.} 
The goal of this research question is to provide an overview of the techniques investigated in the literature to address build failures so that practitioners can find solutions for enhancing software quality and ensuring frequent reliable and successful builds.

\subsubsection{Approach.} 
Similar to RQ\textsubscript{3}, we perform a manual analysis of the papers associated with the topic of build quality to identify the techniques and relevant findings in the studies for addressing build failures. We also examine studies for improving the reliability of build execution, i.e. for \texttt{flaky build detection}. The findings are then summarized and discussed following the sub-categories illustrated in Figure \ref{fig:build_categories}.

\subsubsection{Results.} \textbf{The literature has mainly relied on artificial intelligence techniques to understand, prevent, and repair build failures}, as shown in Table \ref{tab:build_quality_techniques}.
The table shows that machine learning, statistical analysis, natural language processing (NLP), and information retrieval (IR) techniques are the four most widely used techniques, respectively accounting for 40\%, 35\%, 10\%, and 10\% of the 40 examined studies (37 related to \texttt{build quality} and 3 to \texttt{flaky build detection} as depicted in Figure \ref{fig:build_categories}). The remaining studies used data mining (9.4\%) and deep learning (3.1\%) techniques. In the following, we provide more details on the techniques.

\textbf{Existing studies used various statistical analysis and machine learning techniques to investigate the factors causing build failures.} \citet{seo_programmers_2014} used the open coding \cite{corbin_basics_2008} technique to categorize build failures at Google based on historical failed build logs. They found that build failures are mostly due to compiler errors including dependency issues, syntactic errors, and missing repository artifacts. These findings have been further validated by additional studies in industrial \cite{kerzazi_why_2014, rabbani_revisiting_2018} and open source \cite{zhang_large-scale_2019} contexts. \citet{maes-bermejo_revisiting_2022} also investigated the reasons why past builds that were later rerun failed and came to the same conclusion. Further studies \cite{beller_oops_2017, islam_insights_2017, rausch_empirical_2017} used statistical tests (e.g., Shapiro-Wilk Normality test\footnote{\url{https://real-statistics.com/tests-normality-and-symmetry/statistical-tests-normality-symmetry/shapiro-wilk-test}}, %
Wilcoxon Rank Sum test\footnote{\url{https://real-statistics.com/non-parametric-tests/wilcoxon-rank-sum-test}} and effect size measures\footnote{\url{https://real-statistics.com/hypothesis-testing/effect-size}}) to conduct empirical analysis on the build failure factors. In general, they have identified test failures, task complexity, and environment-related issues as the main causes of build failures. Environment-related issues have even been identified as dominant in several studies \cite{ghaleb_studying_2021, zolfagharinia_not_2017, zolfagharinia_study_2019}.

\begin{footnotesize}
\begin{longtable}{p{0.16\linewidth} m{0.17\linewidth} p{0.54\linewidth} p{0.04\linewidth}}
    \caption{Techniques investigated in the literature to improve build quality.}
    \label{tab:build_quality_techniques}\\
    \toprule
        \textbf{Topic} & \textbf{Technique} & \textbf{Description} & \textbf{Ref.} \\
    \midrule
 \endfirsthead

 \multicolumn{4}{c}{Continuation of Table \ref{tab:build_quality_techniques}}\\
    \toprule
        \textbf{Topic} & \textbf{Technique} & \textbf{Description} & \textbf{Ref.}\\
    \midrule
 \endhead

    \bottomrule
 \endfoot

    \bottomrule
 \endlastfoot

        \multirow{46}{\linewidth}{Build Failure \break Analysis (23)} 
                & \multirow{26}{\linewidth}{Statistical Analysis (14)}  & Studied the different categories of compiler errors in 26.6M builds using open coding for log analysis. & \cite{seo_programmers_2014} \\ \cmidrule{3-4}
                & & Studied build failure factors on 3k+ builds collected in an industrial context. & \cite{kerzazi_why_2014} \\ \cmidrule{3-4}  
                & & Examined the impact of testing on build failures using statistical tests. & \cite{beller_oops_2017} \\ \cmidrule{3-4}
                & & Investigated the cause of build failures in Java open-source software (OSS) using an open coding-based tool. & \cite{rausch_empirical_2017} 
                \\ \cmidrule{3-4}
                & & Analyzed build results and their interplay with operating systems and build environments in 30M CPAN builds. & \cite{zolfagharinia_not_2017} \\ \cmidrule{3-4}
                & & Studied build result factors in a 3.6M builds dataset extending TravisTorrent. & \cite{islam_insights_2017} \\ \cmidrule{3-4}
                & & Studied programming errors using Visual Studio Code interaction log data. & \cite{rabbani_revisiting_2018} \\ \cmidrule{3-4}
                & & Studied compiler errors in 6.8M Java builds and examined the 10 most common errors with their fix patterns. & \cite{zhang_large-scale_2019} \\ \cmidrule{3-4}
                & & Studied the root cause of failures in the context of CPAN builds on different environment configurations. & \cite{zolfagharinia_study_2019} \\ \cmidrule{3-4}
                & & Analyzed failure rate and fix time of 857k failed Docker builds in OSS. & \cite{wu_empirical_2020} \\ \cmidrule{3-4}
                & & Examined the errors preventing the build of past snapshots of Java projects. & \cite{maes-bermejo_revisiting_2022}\\ \cmidrule{3-4}
                & & Analyzed the impact of code change types on build failures using the \textsc{SmartSHARK} dataset. & \cite{shimmi_association_2024}\\\cmidrule{3-4}
                & & Studied CI practices associated with successful builds across GitHub OSS. & \cite{wrobel_using_2023}\\  \cmidrule{3-4}
                & & Studied the impact of code ownership of DevOps artifacts on build outcomes using mixed effects logistic regression and linear regression models. & \cite{kola-olawuyi_impact_2024}\\ 
                
                \cmidrule{2-4}
                
                & \multirow{17}{\linewidth}{Machine Learning (9)} & Analyzed build outcome factors in TravisTorrent using clustering models. & \cite{dimitropoulos_continuous_2017} \\ \cmidrule{3-4}
                & &  Studied the factors impacting the build result in TravisTorrent using correlation analysis and feature importances of ML models. & \cite{luo_what_2017} \\ \cmidrule{3-4}
                & &  Identified the most influential features explaining build failures at Atlassian using feature importances of a Logistic Regression model. & \cite{hong_practitioners_2024} \\ \cmidrule{3-4}
                & & Compared the different categories of build errors in open source software versus industrial context using cluster analysis. & \cite{vassallo_tale_2017} \\ \cmidrule{3-4}
                & & Compared the quality of real-world Dockerfiles collected from GitHub projects to a gold set of Dockerfiles using tree association rules. & \cite{henkel_learning_2020} \\ \cmidrule{3-4}
                & & Examined the causal relationship between build failures and code quality factors, by training models using code and test smell datasets. & \cite{barrak_why_2021} \\ \cmidrule{3-4}
                & & Studied the relationship between build duration and build failures using ML. & \cite{ghaleb_studying_2022} \\ \cmidrule{3-4}
                & & Examined the factors influencing build failure correction time using association rules mining. & \cite{silva_what_2023} \\ \cmidrule{3-4}
                & & Studied how contextual project factors influence build outcomes. & \cite{benjamin_study_2023}
                \\ \hline \noalign{\smallskip}

\multirow{8}{\linewidth}{Build Co-change Prediction (4)} & 
        \multirow{8}{\linewidth}{Machine Learning (4)} & Analyzed the factors driving build maintenance during development activity using statistical analysis and association rules. & \cite{mcintosh_empirical_2011} \\ \cmidrule{3-4}
        & & Built Random Forest classifiers to predict and explain the need for build changes using source code and test file changes metrics. & \cite{mcintosh_mining_2014} \\ \cmidrule{3-4}
        & & Predicted build co-change on new projects with small or no historical build data using data from other projects. & \cite{xia_cross-project_2015} \\ \cmidrule{3-4}
        & & Improved the previous build co-change prediction models \cite{mcintosh_mining_2014, xia_cross-project_2015} & \cite{macho_predicting_2016} \\ 
        
        \hline \noalign{\smallskip}

\multirow{19}{\linewidth}{Automatic \break Build Repair (10)} & 

        \multirow{7}{\linewidth}{Information Retrieval (4)} & Summarized failed build logs and mined Stack Overflow (SO) for fixes. & \cite{vassallo_-break_2018} \\ \cmidrule{3-4}
        & & Improved the tool with error localization and more accurate fixes from SO. & \cite{vassallo_every_2020} \\ \cmidrule{3-4}
        & & Surveyed build issue resolution on SO to provide a catalog of build issues. & \cite{lou_understanding_2020} \\ \cmidrule{3-4}
        & & \textit{UniLoc}. Leveraged build logs as search queries and source codes and build scripts as documents to improve fault localization. & \cite{hassan_uniloc_2023} \\

        \cmidrule{2-4}
        
        & \multirow{7}{\linewidth}{Natural Language Processing (4)}
        & Developed three repair strategies for dependency issues and analyzed logs to identify the right strategy. & \cite{macho_automatically_2018} \\ \cmidrule{3-4}
        & & Generated patch candidates to failed Gradle build scripts using log similarity and historical fix patterns. & \cite{hassan_hirebuild_2018} \\ \cmidrule{3-4}
        & & Repaired builds based on build log analysis and historical fix patterns. & \cite{hassan_tackling_2019} \\ \cmidrule{3-4}
        & & Improved SOTA by using current code changes and online resources. & \cite{lou_history-driven_2019} \\ 
        
        \cmidrule{2-4}

        & \multirow{4}{\linewidth}{Deep Learning \& \break LLMs (2)} & Trained a deep neural network to produce the fix to compiler errors using Neural Machine Translation and historical fix patterns. & \cite{mesbah_deepdelta_2019} \\ \cmidrule{3-4}
        & & Fine-tuned the BERT model for sequence classification to automate the classification of build failure logs into fixing categories. & \cite{lee_applying_2024} \\

        \hline \noalign{\smallskip}

\multirow{4}{\linewidth}{Flaky Build \break Detection (3)} & 
        
        \multirow{4}{\linewidth}{Machine Learning (3)}
        & Used job telemetry data with tree-based ML classifiers to detect flaky builds. & \cite{lampel_when_2021} \\ \cmidrule{3-4}
        
        & & Improved flaky build detection models using build logs textual features. & \cite{olewicki_towards_2022} \\ \cmidrule{3-4}
        
        & & Used knowledge graphs of build data to detect and diagnose flaky builds. & \cite{moriconi_automated_2022} \\
        
\end{longtable}
\end{footnotesize}

Numerous studies have also investigated ML techniques to identify the most influential factors of build failures, aiding in the improvement of development and CI practices as well as build failure prediction performance. \citet{dimitropoulos_continuous_2017} used the K-means algorithm for clustering the features in the \textsc{TravisTorrent} dataset, to identify the set of features most correlated with build outcomes. As a result, they determined that features related to large code changes significantly impact the build outcome. Several other studies \cite{luo_what_2017, barrak_why_2021, wrobel_using_2023, silva_what_2023} have come to similar conclusions after conducting exploratory data analysis or modeling build results with ML. Specifically, they found that in open source software (OSS), build failures are mostly explained by the build complexity indicated by a high number of commits or code changes. Further, \citet{silva_what_2023} found that build complexity metrics and project-level characteristics also significantly influence build repair durations. Given the widespread use of container tools in CI, \citet{henkel_learning_2020} assessed the quality of Dockerfiles against best practices using association rule mining and developed a set of rules for improving Dockerfiles.  

Other non-code-related factors of build failures have been identified in OSS projects. For instance, \citet{shimmi_association_2024} investigated the impact of various types of code changes on build failures using discriminant function analysis\footnote{\url{https://real-statistics.com/multivariate-statistics/discriminant-analysis/linear-discriminant-analysis}}, and identified maintenance changes (e.g., bug fixes) as more influential than changes for adding new functionalities. \citet{kola-olawuyi_impact_2024} studied the impact of developers' code ownership of DevOps artifacts on build outcomes and ascertained that successful builds are associated with increased ownership over time and contributions from developers more experienced with DevOps. Contrasting prior studies, \citet{benjamin_enhancing_2024} determined that contextual project-level metrics alone can sufficiently explain build failures. Earlier, \citet{ghaleb_studying_2022} studied the relationship between build failures and duration and highlighted that some actions to improve build duration performance (e.g., caching dependencies) can later lead to build failures (e.g., outdated dependencies), and vice versa. Such a finding underscores the need for more holistic build optimization solutions.

As previous findings were limited to OSS, \citet{vassallo_tale_2017} used K-means clustering to compare build failure reasons in OSS and industrial projects, finding %
different primary causes. In particular, they determined that OSS build failures often stem from unit testing issues, while industrial failures are largely due to release preparation and deployment problems. Recently, \citet{hong_practitioners_2024} analyzed the factors impacting build failures at Atlassian using an LR model and found that the most influential factors are related to the past build performance, such as the ratio of failed builds within the last five builds.

\textbf{Several machine learning techniques have been evaluated to understand and predict when source code changes require build script updates (build co-changes) to prevent build failures.} \citet{mcintosh_empirical_2011} used association rule mining to identify factors driving build script maintenance and trained an RF classifier with code change metrics to predict build co-changes \cite{mcintosh_mining_2014}. They identified a strong relationship between code changes and build script changes, and obtained a prediction accuracy of up to 88\%. Predicting build co-changes requires sufficient historical data about code changes, which are typically limited in newly initiated projects. To address this issue, \citet{xia_cross-project_2015} leveraged the limited training data and non-project-specific features from older projects to train an adaptation of the AdaBoost\footnote{\url{
https://en.wikipedia.org/wiki/AdaBoost}} algorithm. %
\citet{macho_predicting_2016} improved the existing build co-changes prediction models by using detailed code changes and commits categories features to train RF classifiers. This technique outperformed existing models, both for intra-project and cross-project predictions.

\textbf{Various techniques, from mining online resources to deep learning models, have been investigated to generate patch candidates and predict repair strategies for automatic build failure repair}. In the IR techniques, \citet{vassallo_-break_2018} developed a tool that uses a three-step approach to provide a summary of errors and repair hints to help developers understand and repair build failures more quickly. Essentially, the tool (1) extracts the summary and root cause of failures from build log files, then (2) mines the Stack Overflow (SO) platform to find potential repair solutions in the posts, and finally (3) presents to developers the most relevant posts identified using a keyword-based ranking algorithm. \citet{vassallo_every_2020} evaluated the tool as relevant in practice to address failures due to compiler errors. \citet{lou_understanding_2020} have also mined SO to identify the different categories of build errors and how they are fixed in practice.

Repair suggestions that are retrieved from the internet often fail to apply to the specific context of projects. To remedy this, \citet{hassan_hirebuild_2018} used a similarity measure of the build logs with past failed build logs, to identify previous fix patterns used in the projects. From these fix patterns, they generated and evaluated patch candidates. The technique was able to repair 46\% of the studied build failures taken from \textsc{TravisTorrent}.  Similarly, \citet{hassan_tackling_2019} pre-generated build fix templates created from historical code changes of how developers fixed past issues. To repair a failure, they used IR techniques to identify the root cause and location of errors from the build logs, then generated patch candidates using the fix templates. \citet{mesbah_deepdelta_2019} used a deep neural network to generate sets of fixes for compiler errors at Google. The model has been trained with patterns of compiler error resolution, extracted from historical code changes, and encoded into abstract syntax tree\footnote{\url{https://en.wikipedia.org/wiki/abstract_syntax_tree}} (AST) features. %
As a result, the model inferred accurate repairs for 50\% of failures related to dependency and typo issues.

Using historical fix patterns to generate patches is limited by the quality of the old fixes. To address this issue, \citet{macho_automatically_2018} predetermined three repair strategies for dependency-related failures: updating the version, removing the dependency, or adding a repository. In the event of a build failure, the build logs are analyzed to identify the most appropriate repair strategy before it is executed. The technique successfully repaired 54\% of the studied builds. Further, \citet{lou_history-driven_2019} improved the SOTA history-driven build repair technique \cite{hassan_hirebuild_2018}, by relying on the present build code information and online resources to generate the patch candidates. This way, they fixed twice as many build failures as the original technique. 

Most existing build repair techniques focus on error localization within the source code, without considering the build script. To address this shortcoming, \citet{hassan_uniloc_2023} developed \textit{UniLoc}, an IR-based technique that uses build logs as search queries to identify in the source code or build scripts the precise location of errors. An evaluation of this tool on 600 CI failures outperformed SOTA techniques for build error localization. A recent work \cite{lee_applying_2024} using LLMs for automated build repair labeled over 16k Java build error logs and fine-tuned a BERT model for fix strategies classification. The model performed exceptionally well (F1-score: 0.91–1) for categories of fixing strategies with sufficient examples ($\geq$ 14) but struggled with less frequent ones.

\textbf{A limited number of studies have investigated machine learning-based approaches for detecting and diagnosing flaky build failures caused by CI environmental issues.} At Mozilla, \citet{lampel_when_2021} investigated the automated distinction between intermittent (i.e. flaky) build job failures caused by environmental problems and legitimate failures that reveal code issues, using tree-based ML classifiers. These classifiers were trained on telemetry data collected from TaskCluster, the in-house Mozilla CI system. The studied classification models achieved high precision scores (73\% on average) and enabled the analysis of influential features for diagnosing the root causes of flaky job failures.

Telemetry data can be unavailable (or inaccessible) when using public cloud infrastructures to run the builds. To circumvent this need, \citet{olewicki_towards_2022} leveraged historical build logs to detect flaky build jobs at Ubisoft. In particular, they created a vector-based representation of the logs using TF-IDF\footnote{\url{https://en.wikipedia.org/wiki/tf-idf}} %
and leveraged the textual features to train a flaky build classifier with the XGBoost algorithm. This technique achieved almost similar performances (up to 88\% F1-score) as a manual classification, but faster. At Amadeus, \citet{moriconi_automated_2022} evaluated the use of Knowledge Graphs for flaky build detection, which they claimed is better suited for representing heterogeneous build data. For this purpose, they created a labeled dataset of builds using Knowledge Graph Embedding\footnote{\url{https://en.wikipedia.org/wiki/knowledge_graph_embedding}} (KGE) methods, with which they trained an RF classifier that outperformed a baseline TF-IDF-based ML model.

\begin{summary}[RQ\textsubscript{4}]
Several studies have investigated the root causes of build failures to improve code quality and build process reliability. While there is a consensus that build complexity metrics (e.g., high number of commits or code changes) are associated with more failures, other factors such as developer experience and types of changes are also identified. To address build failures, multiple AI and IR techniques have been developed for automatic build repair, build maintenance prediction, and flaky build detection. We suggest future research on the diagnosis and reparation of flaky builds which remains largely underexplored. In addition, we recommend more investigation into the quality of build packaging, particularly in the context of the growing use of container tools. Finally, we suggest more studies in industrial setups, which appear to face unique challenges related to deployment and flakiness.
\end{summary}

\subsection{\rqfive}

\subsubsection{Motivation.} The goal of this research question is to provide an overview of the build metrics that are collected and used in build studies. The findings will be useful to researchers in the creation of more build datasets (as suggested in prior studies \cite{saidani_bf-detector_2021, abdalkareem_machine_2021, hassan_change-aware_2017}), that can be leveraged for future build studies. Our results will also help practitioners identify metrics to collect for implementing ML-based solutions for build optimization.

\subsubsection{Approach.} To answer RQ\textsubscript{5}, we identify the papers presenting a listing (table or enumeration) of the build metrics used in their study. For this purpose, we review the \textit{Data Collection} section, also named \textit{Data Extraction}, \textit{Data Acquisition}, or \textit{Dataset}. As a result, we identified 37 papers highlighted in our replication package. Apart from the paper of \citet{beller_travistorrent_2017} which introduced the \textsc{TravisTorrent} dataset, 12 papers \cite{rausch_empirical_2017, dimitropoulos_continuous_2017, ghaleb_studying_2022, luo_what_2017, barrak_why_2021, islam_insights_2017, silva_what_2023, kola-olawuyi_impact_2024, benjamin_study_2023, hong_practitioners_2024, ghaleb_studying_2021, weeraddana_characterizing_2024} focused on different build data analyses including \texttt{build failure analysis}, \texttt{build duration analysis} and build \texttt{data quality} for ML models, while the remaining 24 papers \cite{hassan_using_2006, saidani_improving_2022, finlay_data_2014, ni_cost-effective_2017, xie_cutting_2018, chen_buildfast_2020, saidani_predicting_2020, hassan_change-aware_2017, xia_could_2017, al-sabbagh_predicting_2022, xia_empirical_2017, jin_cost-efficient_2020, abdalkareem_machine_2021, saidani_detecting_2021, bisong_built_2017, ghaleb_empirical_2019, macho_predicting_2016, mcintosh_mining_2014, xia_cross-project_2015, lampel_when_2021, sun_ravenbuild_2024, wang_commit_2024, benjamin_enhancing_2024, kawalerowicz_continuous_2023} evaluated various ML-based techniques for \texttt{build result prediction}, \texttt{build selection}, \texttt{build duration prediction}, \texttt{build co-change prediction}, and \texttt{falky build detection}. We thoroughly examine the papers, reporting information regarding (1) the description of the metrics used, (2) the categories of these metrics, and (3) the references of the papers that used these metrics. Finally, using the card sorting technique we group the metrics based on the categories identified in the papers.

For synthesis and ease of reuse purposes, we report only on high-level metrics found in each category and that are cited by more than one paper. For example, we present mostly file-level metrics when discussing code-related metrics. However, there are line-level measures that are analogous to some file-level metrics (e.g. number of files/lines added, deleted, and modified) which we have excluded to avoid verbosity, but which have been widely used in the literature \cite{finlay_data_2014, hassan_change-aware_2017, macho_predicting_2016, mcintosh_mining_2014, xia_cross-project_2015}. We also leave out metrics that are too context-specific and therefore can not be easily generalized, like certain metrics that have been created in the context of game software \cite{sun_ravenbuild_2024}.

\begin{small}
\begin{table}
    \caption{List of CI metrics. Each metric is described with its name, its data type, a short description, and a list of study references where it has been used.}
    \label{tab:ci_metrics}
\begin{threeparttable}
    
    \begin{tabular}{>{\raggedright\arraybackslash}m{.4cm} >{\raggedright\arraybackslash}m{2.7cm}  >{\centering\arraybackslash}m{.5cm} m{7.5cm}  >{\scriptsize\raggedright\arraybackslash}m{3cm}}
        \toprule
        \textbf{ID} & \textbf{Metric} & \textbf{DT}\tnote{a} & \textbf{Description} & \small\textbf{Studies} \\
        \midrule
        $B_{1}$ & On default branch & C & Whether the build is triggered on the default branch or not. & \cite{beller_travistorrent_2017, ghaleb_studying_2021, islam_insights_2017, chen_buildfast_2020, wang_commit_2024} \\ \midrule
        
        $B_{2}$& Build type & C & The type of build: push, pull request, merge, or unknown. & \cite{beller_travistorrent_2017, ghaleb_studying_2021, rausch_empirical_2017, dimitropoulos_continuous_2017, ghaleb_studying_2022, luo_what_2017, barrak_why_2021, saidani_improving_2022, saidani_predicting_2020, abdalkareem_machine_2021, saidani_detecting_2021, bisong_built_2017, ghaleb_empirical_2019, kola-olawuyi_impact_2024} \\ \midrule
        
        $B_{3}$ & Time of the day & C & Adjusted hour of the day [0, 23] of the commit that triggered the build (or of the build itself). & \cite{beller_travistorrent_2017, rausch_empirical_2017, ghaleb_studying_2022, luo_what_2017, barrak_why_2021, hassan_using_2006, chen_buildfast_2020, jin_cost-efficient_2020, ghaleb_empirical_2019, benjamin_enhancing_2024, kawalerowicz_continuous_2023, kola-olawuyi_impact_2024} \\ \midrule
        
        $B_{4}$ & Day of the week & C & Adjusted weekday [0, 6] (0 being Monday) of the commit that triggered the build (or of the build itself). & \cite{ghaleb_studying_2021, rausch_empirical_2017, ghaleb_studying_2022, luo_what_2017, barrak_why_2021, hassan_using_2006, chen_buildfast_2020, saidani_predicting_2020, hassan_change-aware_2017, jin_cost-efficient_2020, saidani_detecting_2021, ghaleb_empirical_2019, kawalerowicz_continuous_2023, kola-olawuyi_impact_2024} \\ \midrule
        
        $B_{5}$ & \#Jobs & N & Number of jobs executed in the build. & \cite{beller_travistorrent_2017, ghaleb_studying_2022, bisong_built_2017, benjamin_enhancing_2024} \\

        \midrule\midrule
        
        $B_{6}$ & Build duration & N & Total duration of the last build. & \cite{beller_travistorrent_2017, dimitropoulos_continuous_2017, ghaleb_studying_2022, chen_buildfast_2020, bisong_built_2017, ghaleb_empirical_2019, weeraddana_characterizing_2024, wang_commit_2024, benjamin_enhancing_2024} \\ \midrule

        $B_{7}$ & Build wait time & N & Total wait (or queued) time before executing the build jobs. & \cite{weeraddana_characterizing_2024, benjamin_enhancing_2024} \\ \midrule
        
        $B_{8}$ & Build status & C & Result of the last build: passed, failed, errored, timed out, or canceled. & \cite{beller_travistorrent_2017, ghaleb_studying_2021, rausch_empirical_2017, luo_what_2017, barrak_why_2021, ni_cost-effective_2017, xie_cutting_2018, chen_buildfast_2020, saidani_predicting_2020, hassan_change-aware_2017, hong_practitioners_2024, weeraddana_characterizing_2024, wang_commit_2024, kawalerowicz_continuous_2023, kola-olawuyi_impact_2024} \\ \midrule\midrule
        
        $B_{9}$ & CI lifespan & N & Time difference between the project's last and first builds. & \cite{ghaleb_studying_2021, ghaleb_studying_2022} \\ \midrule

        $B_{10}$ & Time interval & N & Interval of time since the last triggered build (or commit). & \cite{beller_travistorrent_2017, ghaleb_studying_2021, ni_cost-effective_2017, saidani_predicting_2020, jin_cost-efficient_2020, abdalkareem_machine_2021, saidani_detecting_2021, wang_commit_2024, benjamin_enhancing_2024} \\ \midrule
        
        $B_{11}$ & Failure distance & N & The number of builds since the last build failure. & \cite{beller_travistorrent_2017, barrak_why_2021, ni_cost-effective_2017, chen_buildfast_2020, jin_cost-efficient_2020, wang_commit_2024} \\ \midrule
        
        $B_{12}$ & Build performance & N & Ratio of failed builds in the previous builds. & \cite{beller_travistorrent_2017, ghaleb_studying_2021, barrak_why_2021, ni_cost-effective_2017, chen_buildfast_2020, saidani_predicting_2020, jin_cost-efficient_2020, hong_practitioners_2024, wang_commit_2024, kola-olawuyi_impact_2024} \\  \midrule

        $B_{13}$ & Failed builds fix time & N & Time taken to fix a build failure in a specified time interval. & \cite{benjamin_enhancing_2024, silva_what_2023} \\  \midrule\midrule
        
        $B_{14}$ & Job type & C & Type of the build job (e.g., compilation, testing). & \cite{ghaleb_studying_2022, lampel_when_2021} \\ \midrule
        
        $B_{15}$ & Job environment & C & Environment where the build job was run (e.g., JDK version). & \cite{ghaleb_studying_2022, barrak_why_2021, lampel_when_2021} \\ \midrule

        $B_{16}$ & Job logs & N & Execution logs, typically transformed into textual features. & \cite{olewicki_towards_2022, brandt_logchunks_2020} \\ \midrule

        $B_{17}$ & CI Server OS & C & Name and distribution of the OS used to run the build job. & \cite{ghaleb_studying_2022, lampel_when_2021} \\

        \bottomrule
    \end{tabular}
\begin{tablenotes}\footnotesize
\item[a] \textbf{Data Type (DT)}: Categorical (C) – Numeric (N).
\end{tablenotes}
\end{threeparttable}
\end{table}
\end{small}

\subsubsection{Results.} \textbf{We identified four main categories of metrics used to conduct research on the build, the most prevalent of which are code metrics.} Tables \ref{tab:ci_metrics}, \ref{tab:project_metrics}, \ref{tab:developer_metrics}, \ref{tab:code_metrics}
 present the four main categories of build metrics used in the literature, respectively: CI metrics, project metrics, developer metrics, and code metrics. Code metrics are the most widely used category of build metrics identified. As shown in the tables, code metrics have been identified in 36 (97.3\%) of the 37 studies examined. Meanwhile, CI metrics, developer metrics, and project metrics were identified in 30 (81.1\%), 26 (70.3\%), and 21 (56.8\%) studies respectively. In the following, we present in more detail each category of build metrics.

\textbf{CI metrics are related to the current build, as well as the historical build activity and performance of projects}. Table \ref{tab:ci_metrics} summarizes the CI metrics identified in the literature. The raw data for calculating these metrics can be collected mainly from the CI system\footnote{An example of data source for the CI metrics on the GitLab CI system is the \href{https://docs.gitlab.com/ee/api/pipelines.html}{GitLab Pipelines API}} and the VCS. The CI metrics can be grouped into four sub-categories. First, metrics that pertain to the characteristics of the build before its execution ($B_{1}$ --- $B_{5}$). Second, metrics related to a single build execution performance and outcome ($B_{6}$ --- $B_{8}$). Third, metrics that provide general insights into the historical performance of the CI process and are calculated using past build data ($B_{9}$ --- $B_{13}$). Finally, the fourth sub-category includes job-specific metrics such as execution logs and details about the environment used to run the build jobs ($B_{14}$ --- $B_{17}$). These job-specific metrics are the least used in the literature, found in only five studies.

Different synonyms and proxies of CI metrics are used in the literature. For instance, \citet{ghaleb_studying_2021} referred to the \texttt{CI lifespan} ($B_{9}$) as the \texttt{build age}, and \citet{ni_cost-effective_2017} used the name \texttt{project recent} to indicate the same measure as the \texttt{build performance} ($B_{12}$). Examples of proxy metrics include \texttt{day/night} and \texttt{weekday/weekend} metrics which have been used instead of the \texttt{time of the day} ($B_{3}$) and \texttt{day of the week} ($B_{4}$) \cite{ghaleb_studying_2022}. Additionally, in place of the \texttt{build type} ($B_{2}$), other studies \cite{ghaleb_studying_2021, beller_travistorrent_2017} used a boolean proxy metric \texttt{is pull request} indicating whether the build is triggered from a pull request or not.
Further, instead of the \texttt{build performance} ($B_{12}$), \citet{rausch_empirical_2017} used a proxy metric \texttt{build climate} defined as the failure rate in the n (e.g., 5) recent builds. 

\textbf{Project metrics provide high-level information about the characteristics and maturity of the projects}. These metrics include the age of the project ($P_{3}$), its size in lines of code ($P_{4}$), and other general information on the project such as the dominant programming language ($P_{1}$) and the size of the team contributing to the project ($P_{2}$). Table \ref{tab:project_metrics} summarizes the project metrics identified. These metrics are typically extracted from the VCS.

Several project metrics have been defined differently across the studies. For instance, while some studies \cite{beller_travistorrent_2017, chen_buildfast_2020} determined the \texttt{team size} ($P_{2}$) based on (the authors of) contributions made during the three months preceding the last build of the project, another study \cite{islam_insights_2017} considered the contributions made in the whole project history. In addition, \citet{ghaleb_studying_2022} defined the \texttt{project age} ($P_{3}$) as the number of days between the last build and the project creation date, while for \citet{jin_cost-efficient_2020}, it is the time difference between the last and the first build of the project also known as the CI lifespan \cite{benjamin_study_2023}. Similarly, prior studies \cite{ghaleb_studying_2022, jin_cost-efficient_2020} defined \texttt{test density} ($P_{5}$) as the median number of test cases or test lines (per 1,000 source lines of code), while other studies \cite{dimitropoulos_continuous_2017, beller_travistorrent_2017, saidani_predicting_2020} considered the total number of test cases or test lines per 1,000 SLOC.

\begin{small}
\begin{table}
\caption{List of project metrics}
\label{tab:project_metrics}
\begin{tabular}{>{\raggedright\arraybackslash}m{.4cm} >{\raggedright\arraybackslash}m{2.7cm}  >{\centering\arraybackslash}m{.5cm}  m{7.5cm}  >{\scriptsize\raggedright\arraybackslash}m{3cm}}
\toprule
\textbf{ID} & \textbf{Metric} & \textbf{DT} & \textbf{Description} &  \small\textbf{Studies} \\
\midrule
$P_{1}$ & Language & C & The main programming language of the project. & \cite{beller_travistorrent_2017, ghaleb_studying_2021, dimitropoulos_continuous_2017, ghaleb_studying_2022, luo_what_2017, ghaleb_empirical_2019, benjamin_enhancing_2024, benjamin_study_2023} \\ \midrule

$P_{2}$ & Team size & N & The number of contributors in the team working on the project. & \cite{beller_travistorrent_2017, ghaleb_studying_2021, dimitropoulos_continuous_2017, ghaleb_studying_2022, luo_what_2017, islam_insights_2017, saidani_improving_2022, chen_buildfast_2020, saidani_predicting_2020, hassan_change-aware_2017, xia_could_2017, al-sabbagh_predicting_2022, jin_cost-efficient_2020, bisong_built_2017, ghaleb_empirical_2019, wang_commit_2024, silva_what_2023, benjamin_study_2023}\\ 

\midrule

$P_{3}$ & Project age & N & Time difference (in days) between the last executed build and the project creation date. & \cite{ghaleb_studying_2022, jin_cost-efficient_2020, benjamin_enhancing_2024, silva_what_2023} \\ \midrule

$P_{4}$ & Project size & N & Size of the project in Source Lines Of Codes (SLOC). & \cite{ghaleb_studying_2022, jin_cost-efficient_2020, bisong_built_2017, ghaleb_empirical_2019, mcintosh_mining_2014, benjamin_enhancing_2024, silva_what_2023, benjamin_study_2023} \\ \midrule

$P_{5}$ & Test density & N & Median number of test cases or test lines per 1k SLOC. & \cite{dimitropoulos_continuous_2017, ghaleb_studying_2022, luo_what_2017, saidani_improving_2022, saidani_predicting_2020, xia_could_2017, xia_empirical_2017, jin_cost-efficient_2020, bisong_built_2017, ghaleb_empirical_2019, benjamin_study_2023} \\

\bottomrule
\end{tabular}
\end{table}
\end{small}

\begin{small}
    \begin{table}
    \caption{List of developer metrics}
    \label{tab:developer_metrics}
        \begin{tabular}{>{\raggedright\arraybackslash}m{.4cm} >{\raggedright\arraybackslash}m{2.7cm}  >{\centering\arraybackslash}m{.5cm}  m{7.5cm}  >{\scriptsize\raggedright\arraybackslash}m{3cm}}
            \toprule
            \textbf{ID} & \textbf{Metric} & \textbf{DT} & \textbf{Description} & \small\textbf{Studies} \\
            \midrule
            
            $D_{1}$ & \#Contributors & N & Number of distinct authors of the code changes since the last build. & \cite{beller_travistorrent_2017, rausch_empirical_2017, dimitropoulos_continuous_2017, barrak_why_2021, hassan_using_2006, saidani_improving_2022, xie_cutting_2018, chen_buildfast_2020, saidani_predicting_2020, xia_empirical_2017, abdalkareem_machine_2021, saidani_detecting_2021, bisong_built_2017, hong_practitioners_2024, sun_ravenbuild_2024, kawalerowicz_continuous_2023, kola-olawuyi_impact_2024} \\ \midrule

            $D_{2}$ & Core developer & C & Whether the developer who triggered the build has committed at least one in the last 3 months. & \cite{ghaleb_studying_2021, dimitropoulos_continuous_2017, ghaleb_studying_2022, luo_what_2017, barrak_why_2021, saidani_improving_2022, chen_buildfast_2020, saidani_predicting_2020, xia_could_2017, xia_empirical_2017, bisong_built_2017, ghaleb_empirical_2019, wang_commit_2024, benjamin_enhancing_2024} \\ \midrule
            
            $D_{3}$ & Dev. experience & N &  Number of days (or commits) the developer who authored the build has been contributing to the project. & \cite{rausch_empirical_2017, ghaleb_studying_2022, barrak_why_2021, hassan_using_2006, ni_cost-effective_2017, saidani_predicting_2020, abdalkareem_machine_2021, saidani_detecting_2021, ghaleb_empirical_2019, benjamin_enhancing_2024, silva_what_2023, kola-olawuyi_impact_2024} \\ \midrule

            $D_{4}$ & Dev. failure rate & N &  Failure rate of the builds authored by the developer in the project history. & \cite{saidani_predicting_2020, silva_what_2023, kola-olawuyi_impact_2024} \\ 
            
            \bottomrule
        \end{tabular}
    \end{table}
\end{small}

\textbf{Developer metrics provide information regarding the developers that authored the triggered build}. Table \ref{tab:developer_metrics} summarizes the developer metrics identified, that can be collected from the VCS. These metrics include the number of contributors ($D_{1}$), and the role ($D_{2}$) and experience of the developer who triggered the build with contributions ($D_{3}$) and with failures ($D_{4}$).

Conflicting perspectives on the calculation of specific developer metrics have also been observed. In some studies \cite{ghaleb_studying_2022, ghaleb_empirical_2019}, the \texttt{developer experience} ($D_{3}$) is measured either as the number of days the developer has been contributing to the project or as the number of commits the developer authored in the project to which the build belongs. Moreover, another study \cite{saidani_predicting_2020} argued that the same metric -- $D_{3}$ -- can be measured as the average number of builds authored by the developer or as the build failure rate associated with the developer (i.e., $D_{4}$). 

\begin{small}
\begin{table}
\caption{List of code metrics}
\label{tab:code_metrics}
\begin{tabular}{>{\raggedright\arraybackslash}m{.4cm} >{\raggedright\arraybackslash}m{2.5cm}  >{\centering\arraybackslash}m{.4cm}  m{7.5cm}  >{\scriptsize\raggedright\arraybackslash}m{3.3cm}}
\toprule
\textbf{ID} & \textbf{Metric} & \textbf{DT} & \textbf{Description} & \small\textbf{Studies} \\
\midrule

$C_{1}$ & Commits & N & Number of unique commits pushed since the last build. Also known as the number of revisions \cite{kawalerowicz_continuous_2023}. & \cite{beller_travistorrent_2017, ghaleb_studying_2021, rausch_empirical_2017, ghaleb_studying_2022, luo_what_2017, barrak_why_2021, islam_insights_2017, saidani_improving_2022, ni_cost-effective_2017, xie_cutting_2018, chen_buildfast_2020, saidani_predicting_2020, xia_could_2017, al-sabbagh_predicting_2022, xia_empirical_2017, jin_cost-efficient_2020, abdalkareem_machine_2021, saidani_detecting_2021, bisong_built_2017, ghaleb_empirical_2019, wang_commit_2024, benjamin_enhancing_2024, kawalerowicz_continuous_2023, silva_what_2023, kola-olawuyi_impact_2024} \\ \midrule

$C_{2}$ & Files changed & N & Number of files/lines changed by the commits in the build. & \cite{beller_travistorrent_2017, ghaleb_studying_2021, rausch_empirical_2017, dimitropoulos_continuous_2017, ghaleb_studying_2022, luo_what_2017, barrak_why_2021, islam_insights_2017, hassan_using_2006, saidani_improving_2022, ni_cost-effective_2017, xie_cutting_2018, chen_buildfast_2020, saidani_predicting_2020, hassan_change-aware_2017, xia_could_2017, al-sabbagh_predicting_2022, xia_empirical_2017, saidani_detecting_2021, bisong_built_2017, ghaleb_empirical_2019, macho_predicting_2016, mcintosh_mining_2014, xia_cross-project_2015, hong_practitioners_2024, weeraddana_characterizing_2024, kawalerowicz_continuous_2023, silva_what_2023, kola-olawuyi_impact_2024} \\ \midrule \midrule

$C_{3}$ & Files deleted & N & Number of files/lines deleted by the commits in the build. & \cite{beller_travistorrent_2017, dimitropoulos_continuous_2017, luo_what_2017, saidani_improving_2022, chen_buildfast_2020, saidani_predicting_2020, hassan_change-aware_2017, xia_could_2017, al-sabbagh_predicting_2022, xia_empirical_2017, bisong_built_2017, ghaleb_empirical_2019, macho_predicting_2016, mcintosh_mining_2014, xia_cross-project_2015, weeraddana_characterizing_2024, wang_commit_2024}  \\ \midrule

$C_{4}$ & Files added & N & Number of files/lines added by the commits in the build. & \cite{beller_travistorrent_2017, dimitropoulos_continuous_2017, luo_what_2017, saidani_improving_2022, ni_cost-effective_2017, chen_buildfast_2020, saidani_predicting_2020, hassan_change-aware_2017, xia_could_2017, al-sabbagh_predicting_2022, xia_empirical_2017, bisong_built_2017, ghaleb_empirical_2019, macho_predicting_2016, mcintosh_mining_2014, xia_cross-project_2015, weeraddana_characterizing_2024, wang_commit_2024} \\ \midrule

$C_{5}$ & Tests added & N & Number of added test cases & \cite{beller_travistorrent_2017, ghaleb_studying_2021, dimitropoulos_continuous_2017, luo_what_2017, saidani_improving_2022, saidani_predicting_2020, xia_could_2017, xia_empirical_2017, bisong_built_2017, ghaleb_empirical_2019} \\ \midrule

$C_{6}$ & Tests deleted & N & Number of deleted test cases & \cite{beller_travistorrent_2017, ghaleb_studying_2021, dimitropoulos_continuous_2017, luo_what_2017, saidani_improving_2022, saidani_predicting_2020, xia_could_2017, xia_empirical_2017, bisong_built_2017, ghaleb_empirical_2019} \\ \midrule

$C_{7}$ & Imports added & N & Number of import statements added & \cite{finlay_data_2014, wang_commit_2024} \\ \midrule

$C_{8}$ & Classes changed & N & Number of classes changed, added, or deleted & \cite{finlay_data_2014, wang_commit_2024} \\ \midrule\midrule

$C_{9}$ & Source churn & N & Lines of source (src.) code changed by the commits in the build. Also referred to as the commit length \cite{benjamin_enhancing_2024}. & \cite{beller_travistorrent_2017, ghaleb_studying_2021, dimitropoulos_continuous_2017, ghaleb_studying_2022, luo_what_2017, islam_insights_2017, saidani_improving_2022, finlay_data_2014, xie_cutting_2018, chen_buildfast_2020, saidani_predicting_2020, hassan_change-aware_2017, xia_could_2017, al-sabbagh_predicting_2022, xia_empirical_2017, jin_cost-efficient_2020, abdalkareem_which_2021, bisong_built_2017, ghaleb_empirical_2019, xia_cross-project_2015, weeraddana_characterizing_2024, wang_commit_2024, benjamin_enhancing_2024} \\ \midrule

$C_{10}$ & Test churn & N & Lines of test code changed by the commits in the build. & \cite{beller_travistorrent_2017, ghaleb_studying_2021, dimitropoulos_continuous_2017, ghaleb_studying_2022, luo_what_2017, islam_insights_2017, saidani_improving_2022, chen_buildfast_2020, saidani_predicting_2020, hassan_change-aware_2017, xia_could_2017, xia_empirical_2017, jin_cost-efficient_2020, bisong_built_2017, ghaleb_empirical_2019, xia_cross-project_2015, benjamin_enhancing_2024} \\ \midrule

$C_{11}$ & Src. files changed & N & Number of source files changed by the commits in the build. & \cite{beller_travistorrent_2017, ghaleb_studying_2021, dimitropoulos_continuous_2017, ghaleb_studying_2022, luo_what_2017, saidani_improving_2022, xie_cutting_2018, chen_buildfast_2020, saidani_predicting_2020, xia_could_2017, al-sabbagh_predicting_2022, xia_empirical_2017, jin_cost-efficient_2020, bisong_built_2017, ghaleb_empirical_2019, xia_cross-project_2015, wang_commit_2024, benjamin_enhancing_2024}\\ \midrule

$C_{12}$ & Doc. files changed & N & Number of documentation files changed by the commits in the build. & \cite{beller_travistorrent_2017, ghaleb_studying_2021, dimitropoulos_continuous_2017, ghaleb_studying_2022, luo_what_2017, saidani_improving_2022, chen_buildfast_2020, saidani_predicting_2020, hassan_change-aware_2017, xia_could_2017, al-sabbagh_predicting_2022, saidani_detecting_2021, bisong_built_2017, ghaleb_empirical_2019, macho_predicting_2016} \\ \midrule

$C_{13}$ & Config. files changed & N & Number of configuration (e.g., \texttt{.xml} and \texttt{.yml}) files modified by the commits in the
build. & \cite{ghaleb_studying_2021, ghaleb_studying_2022, chen_buildfast_2020, saidani_predicting_2020, ghaleb_empirical_2019, benjamin_enhancing_2024} \\ \midrule

$C_{14}$ & Other files changed & N & Number of other files changed by the commits in the build. & \cite{beller_travistorrent_2017, ghaleb_studying_2021, dimitropoulos_continuous_2017, ghaleb_studying_2022, luo_what_2017, saidani_improving_2022, saidani_predicting_2020, hassan_change-aware_2017, xia_could_2017, al-sabbagh_predicting_2022, abdalkareem_machine_2021, bisong_built_2017, ghaleb_empirical_2019, sun_ravenbuild_2024} \\

\bottomrule
\end{tabular}
\end{table}
\end{small}

\textbf{Code metrics encompass metrics related to the code changes associated with the triggered build}. Table \ref{tab:code_metrics} summarizes the identified code metrics. These metrics measure the number of code changes made overall in the project ($C_{1}$, $C_{2}$), the number of different types of changes, ($C_{3}$ --- $C_{8}$), and the number of changes affecting specific types of files ($C_{9}$ --- $C_{14}$). The code metrics can be collected from the VCS and using static code analysis tools such as SciTools Understand\footnote{https://scitools.com/}. Depending on the context, $C_{1}$ and $C_{2}$ can also be considered as pull request (PR) metrics (e.g., in \cite{hong_practitioners_2024}) given that they can both reflect the changes associated with the build's PR.

The code metrics are no less subject to variation in definitions and calculation methods. As evidence, \citet{beller_travistorrent_2017} considered the \texttt{tests added} ($C_{5}$) as the number of lines of testing code added since the last build, while for \citet{ghaleb_studying_2021}, $C_{5}$ is the number of test cases added since the last build. Although these two definitions can be regarded as proxies, in reality, they give indications that are different and should not be treated in the same way. For instance, a single test case can be written in the same number of lines as several test cases. Hence, different definitions can lead to confusion in the reuse and interpretation of the metrics. We invite future studies to evaluate the quality of these different types of metrics, to clarify which are the most useful.

\begin{summary}[RQ\textsubscript{5}]
The identified build metrics are grouped into four main categories: CI metrics, project metrics, developer metrics, and code metrics, with code metrics being the most commonly used. The build metrics can be collected from different sources, including APIs exposed by the VCS and the CI system, and static code analysis tools. In the literature, there are different definitions and proxies for build metrics, which can be misleading. We invite researchers to evaluate the different variations of build metrics and create a taxonomy of clearly defined metrics to support future build-related studies.
\end{summary}

\subsection{\rqsix}

\subsubsection{Motivation.} RQ\textsubscript{1} shows that build studies are conducted using different build datasets. This research question aims to provide a benchmark of publicly accessible build datasets that are available for build studies. A comprehensive review of the build datasets will facilitate for researchers, the use and improvement of these datasets for future studies.

\subsubsection{Approach.} Firstly, we read the titles of the 97 papers to identify those that explicitly present build datasets. We found four \texttt{build datasets} papers as discussed in RQ\textsubscript{1}. Furthermore, to identify other datasets made publicly available in different study contexts, we examine the \textit{Data Collection} sections (also named \textit{Data Extraction}, or \textit{Datasets}) of each of the remaining 93 papers. This examination resulted in 4 additional papers, each containing a different dataset. Finally, we review the content of each relevant paper to identify the properties of the dataset (i.e., size, data format, CI system, programming languages, etc.), and the research topics for which the dataset has been created, as mentioned in the abstract or introduction.

Other datasets have been identified through the studies but were not publicly available such as the dataset from Google \cite{seo_programmers_2014}, IBM Jazz \cite{finlay_data_2014}, and CPAN \cite{zolfagharinia_study_2019}. We have not found any indication in the papers that these datasets have been made publicly accessible, nor any information on how they were created.

\begin{small}
\begin{table}
\caption{List of publicly available build datasets. It shows the name of the dataset, its size, the associated CI system, and a short description including the data format, the programming languages covered, and the original purpose of the dataset. The dataset name is clickable and is a link to its DOI or download page.}
\label{tab:datasets}
\begin{tabular}{>{\raggedright\arraybackslash}m{2.8cm} >{\raggedleft\arraybackslash}m{1.3cm} >{\centering\arraybackslash}m{2cm} m{5.2cm} >{\centering\arraybackslash}m{.6cm} >{\centering\arraybackslash}m{.8cm}}

\toprule
\textbf{Name} & \textbf{\# Builds} & \textbf{CI Service} & \textbf{Description} & \textbf{Year} & \textbf{Paper}  \\
\midrule

\href{https://doi.org/10.6084/m9.figshare.19314170.v1}{\textsc{TravisTorrent}} & 2,640,825 & Travis CI &  A dataset of .csv and log files collected from 1,300 Java and Ruby projects using Travis CI; for build failure analysis. & 2017 & \cite{beller_travistorrent_2017} \\ \midrule

\href{https://doi.org/10.5281/zenodo.3539803}{\textsc{CI Compiler Errors}} & 6,854,271 & Travis CI & 3.8TB of .json and log files created from builds in 3,799 Java projects using Travis CI. Created for build failure analysis. & 2019 & \cite{zhang_large-scale_2019} \\ \midrule

\href{https://doi.org/10.5281/zenodo.3632351}{\textsc{LogChunks}} & 797 & Travis CI & Collection of labeled log files collected from 80 projects (29 prog. languages) using Travis CI, for build log analysis. & 2020 & \cite{brandt_logchunks_2020} \\ \midrule

\href{https://doi.org/10.17605/OSF.IO/UMK3W}{\textsc{Local Java Builds}} & 14,497 & N/A & A dataset of .csv and log files collected local build results of 20,000 projects written in Java. Created for build failure analysis. & 2020 & \cite{sulir_large-scale_2020} \\ \midrule

\href{https://doi.org/10.5281/zenodo.4682056}{\textsc{CIBench}} & 82,427 & Travis CI &  Extended TravisTorrent using builds from 100 projects, to evaluate build selection and prioritization techniques. & 2020 & \cite{jin_cibench_2021} \\ \midrule

\href{https://colab.research.google.com/drive/1Eso1Cc380yIt79hrmR25yjEYd-SJOt5b}{\textsc{Mozilla CI Jobs}} & 1,997,075 &  Taskcluster &  CI job telemetry data stored in SQL database to support flaky builds detection. & 2021 & \cite{lampel_when_2021} \\ \midrule

\href{https://doi.org/10.25625/6WAFNG}{\textsc{SmartSHARK}} & 89,160 &  Travis CI &  740 GB MongoDB data from VCS, CI service, and issue tracking across 77 Java projects for software evolution analysis. & 2021 & \cite{shimmi_association_2024} \\ \midrule

\href{https://doi.org/10.5281/zenodo.11192753}{\textsc{Deps Updates Builds}} & 10,412 &  GitHub Actions & 3.3GB of build data related to dependency updates commits of Javascript projects, to analyze unused dependencies' wastes. & 2023 & \cite{weeraddana_dependency-induced_2024} \\

\bottomrule
\end{tabular}

\end{table}
\end{small}

\subsubsection{Results.} \textbf{A total of eight publicly accessible build datasets have been identified in the literature, to support various build research topics, with build failure analysis being the most prominent}. Table \ref{tab:datasets} presents the publicly available build datasets identified. 4 of the 8 datasets have been the exclusive topic of papers \cite{beller_travistorrent_2017, brandt_logchunks_2020, sulir_large-scale_2020, jin_cibench_2021}, while the remaining datasets have been made available as part of studies in different contexts \cite{zhang_large-scale_2019, lampel_when_2021, shimmi_association_2024, weeraddana_dependency-induced_2024}. The \textsc{TravisTorrent} \cite{beller_travistorrent_2017} and \textsc{CI Compiler Errors} \cite{zhang_large-scale_2019} datasets are outstanding for the included number of builds, containing metadata and log files collected across 2.6 and 6.8 million builds, respectively. In addition, the \textsc{Mozilla CI Jobs} \cite{lampel_when_2021} dataset contains telemetry data of nearly 2 million CI jobs. The rest of the datasets \cite{brandt_logchunks_2020, sulir_large-scale_2020, jin_cibench_2021, shimmi_association_2024, weeraddana_dependency-induced_2024} cover a more limited number of builds, ranging from 797 to 89k.

The publicly available build datasets can be leveraged for various build topics. \textsc{TravisTorrent} has been widely used in the literature for research on topics including build failures and duration analysis in \cite{bisong_built_2017, jin_what_2021} for example, build result prediction \cite{saidani_toward_2021, ni_acona_2018}, build selection \cite{abdalkareem_which_2021, jin_reducing_2021}, and automated build repair \cite{hassan_tackling_2019}. The more recent \textsc{SmartSHARK} dataset was developed to support a wide range of research topics, including those explored using \textsc{TravisTorrent}, although it has been used in only one build-related paper so far. 
\textsc{CIBench}, as an extension of \textsc{TravisTorrent}, can also be used for similar research goals. Nevertheless, the authors created this dataset to evaluate and compare build selection and prioritization techniques under the same settings \cite{jin_what_2021}. The log dataset \textsc{LogChunks} has been created to support build log analysis studies including the evaluation of log-based information retrieval and log classification techniques. In contrast, the primary purpose of the \textsc{Local Java Builds} and \textsc{CI Compiler Errors} datasets is to support studies on build failure analysis in Java projects. In addition, the authors proposed various research directions using these datasets, including the exploration of additional root causes of build failures (such as environmental issues) and the evaluation of build failure prediction and automated build repair techniques. Finally, the \textsc{Mozilla CI Jobs} dataset was introduced for studies on detecting flaky builds using telemetry data, while the \textsc{Deps Updates Builds} dataset was intended for future research addressing unused dependencies' waste.

\textbf{Most of the existing build datasets focused on projects written in Java and using Travis CI}. As shown in Table \ref{tab:datasets}, out the eight datasets, five (62.5\%) \cite{beller_travistorrent_2017, zhang_large-scale_2019, brandt_logchunks_2020, jin_cibench_2021, shimmi_association_2024} are created from projects using Travis CI. Also, five (62.5\%) datasets \cite{beller_travistorrent_2017, zhang_large-scale_2019, sulir_large-scale_2020, jin_cibench_2021, shimmi_association_2024} are created from projects mainly written in Java. In particular, the \textsc{Local Java Builds} and \textsc{CI Compiler Errors} datasets contain only Java builds. As a result, the vast majority of existing build studies have been carried out in the context of the Java programming language and the Travis CI system.

Several studies have highlighted the limitations of studying builds in specific contexts, suggesting the need for more data related to other CI platforms and programming languages. Many important aspects of the build (e.g., configuration, environment) are tightly coupled to the build tool and CI system used. Hence, studying builds in the context of a limited number of these tools may hinder the generalization of results and limit their applicability at a broader scale \cite{mcintosh_mining_2014, abdalkareem_which_2021, saidani_bf-detector_2021, lampel_when_2021}. 
These limitations are particularly significant because Travis CI, extensively studied in the literature, has been replaced by GitHub Actions as the most commonly used CI tool \cite{golzadeh_rise_2022}. Additional studies outlined the difficulty of adopting existing techniques due to programming language specifics \cite{abdalkareem_which_2021, olewicki_towards_2022}. In particular, \citet{zolfagharinia_study_2019} argued that the differing specific requirements of programming languages  (such as runtime environments and packages) make it challenging to reuse CI-improving techniques across languages. Also, \citet{mcintosh_mining_2014} noted a performance difference between Java and web-based projects when predicting build co-changes, due to the peculiarities of the build technologies used in each context.

\begin{summary}[RQ\textsubscript{6}]
There are 8 publicly available build datasets in the literature that can be leveraged for research in topics such as build failure analysis, build prediction, build reparation, and flaky build detection. However, the majority of these build data are limited to only projects using Travis CI and written in the Java programming language. Therefore, we encourage the creation of build datasets that cover novel industry standard CI tools and other widely used programming languages.
\end{summary}

\section{Conclusion}
\label{sec:conclusion}

In this paper, we conduct a systematic literature review of 97 papers to better understand the existing practices for build optimization and guide future studies on CI build. More than three-quarters (88.7\%) of the papers were published after 2017, following the release of the \textsc{TravisTorrent} dataset, which was used to carry almost half (41.2\%) of the build studies. Further, the SLR shows that 95.8\% of the build studies have leveraged different build datasets, out of which we have identified 8 publicly available that can be used for future build studies.

Our analysis of the 97 papers shows that build studies have focused on two main challenges: long build times and build failures. Indeed, these two challenges have been associated with significant infrastructure and human (reduced productivity) costs, estimated in millions of dollars in large organizations \cite{zhang_buildsonic_2022}. Besides the papers focusing on these challenges, the remaining papers fall into two main categories: general analysis studies on the use and quality of CI systems, and studies on the datasets used to conduct build studies.

To optimize the build and mitigate its costs, existing studies have developed several techniques and approaches, most of which are based on ML algorithms. For example, ML-based build selection techniques have been developed to minimize execution costs, while predicting build outcomes and durations has proven effective in reducing developer waiting times. As for build failures, existing studies have generally focused on root cause analysis, build maintenance prediction, and automatic build repair. In addition, recent efforts have focused on detecting flaky builds, most often caused by instabilities within the CI system, such as memory errors.

This SLR highlighted several research gaps that warrant attention in future studies. Notably, there is a limited number of studies conducted in industrial settings and within the context of modern CI tools, such as GitHub Actions and GitLab CI, which limits the adoption of existing techniques by practitioners. Expanding the availability of build datasets for these tools would greatly benefit the research community. In addition, we recommend more studies on builds in containerized environments, which have become the industry standard for running CI, and the development of better CI-improving techniques that are still underexplored. Lastly, while recent work has shifted focus toward flaky build failures (beyond flaky tests), research on diagnosing and repairing these environment-related build failures remains scarce and requires further investigation.

\begin{acks}
We acknowledge the support of the Natural Sciences and Engineering Research Council of Canada (NSERC), ALLRP/576653-2022. This work was also supported by Mitacs through the Mitacs Accelerate program.
\end{acks}

\bibliographystyle{ACM-Reference-Format}
\bibliography{bibliography/main}

\appendix

\end{document}